\documentclass[10pt,final,doubleside]{IEEEtran}

\usepackage{pifont,bm,multicol,amsfonts,amsmath,color,amssymb,graphicx,balance, epsfig,cite,psfrag,subfigure,algorithm,enumerate,stfloats,algorithm,algorithmic,epstopdf}

\newtheorem{theorem}{\underline{Theorem}}

\newtheorem{remark}{\underline{Remark}}

\begin{document}

\title{Interference-Alignment and Soft-Space-Reuse  Based Cooperative Transmission   for Multi-cell Massive MIMO Networks}
\author{Jianpeng Ma, { \emph{Student Member, IEEE,}} Shun Zhang, { \emph{Member, IEEE,}} \\Hongyan Li, \emph{Member, IEEE,} Nan Zhao, \emph{Senior Member, IEEE,} \\and Victor C.M. Leung, \emph{Fellow, IEEE}

\thanks{J. Ma, S. Zhang, and H. Li  are with the State Key Laboratory of Integrated Services Networks,
Xidian University, Xi'an 710071, P. R. China. $\big($Email:
jpmaxdu@gmail.com; zhangshunsdu@gmail.com; hyli@mail.xidian.edu.cn$\big)$.}
\thanks{N. Zhao is with the School of Inform. and Commun. Eng., Dalian University of Technology, Dalian, Liaoning, P. R. China (email: zhaonan@dlut.edu.cn).}
\thanks{V.C.M. Leung is with the Department of Electrical and Computer Engineering,
the University of British Columbia, Vancouver, BC, V6T 1Z4, Canada (email: vleung@ece.ubc.ca).}}


{}
\maketitle

\begin{abstract}
 \textcolor{blue}{As a revolutionary wireless transmission strategy,
 	interference alignment (IA) can improve the capacity of the
 	cell-edge users. However, the acquisition of the global channel
 	state information (CSI) for IA leads to unacceptable overhead
 	in the massive MIMO systems.} To tackle this problem, in this paper, we propose an IA and soft-space-reuse (IA-SSR) based cooperative transmission scheme under the two-stage precoding framework. Specifically, {the cell-center and the cell-edge users are separately treated} to fully exploit the spatial degrees of freedoms (DoF). Then, the optimal power allocation policy is developed  to maximize the sum-capacity of the network. Next, a low-cost channel estimator is designed for the proposed IA-SSR framework. \textcolor{blue}{Some practical issues in IA-SSR implementation are also discussed.} Finally, plenty of numerical results are presented to show the efficiency of the proposed algorithm.

\end{abstract}


\IEEEpeerreviewmaketitle

\section{Introduction}

Due to its significant improvement in spectral and power efficiency, the massive multiple-input multiple-output (MIMO) system has been widely considered as a promising technology for the 5th generation (5G) cellular systems    \cite{efficiency2,Massive_in_5G_1,Massive_in_5G_2}.
In order to implement the downlink precoding and the uplink detection, the
 base stations (BSs) of the massive MIMO networks should acquire the accurate channel state information (CSI).

 In  the
 time-division duplex (TDD) systems, the CSI at  the
 BS sides can be obtained through  the  uplink training  with aid of the  uplink-downlink reciprocity.
 Under this  scenario, the length  of  the
 training is proportional to the total number of  the
 user antennas \cite{TDD-massive,feedback}.
 However, in  the
 frequency-division duplex (FDD) systems, the uplink-downlink reciprocity does not exist.
 \textcolor[rgb]{0.00,0.00,1.00}{Then, the CSI at the BS side can only be achieved through three steps, i.e., the downlink training, the channel estimation at the user side, and the CSI feedback.}
 {Correspondingly,
the amount of both the training symbols and the feedback CSI are in scale with the number of the antennas at the BSs,} which will lead to unacceptable overhead \cite{training_fdd}, \cite{training_fdd2}.

To overcome this  {bottleneck},  {a two-stage precoding scheme
called ``joint spatial division and multiplexing  (JSDM)''  was proposed} in \cite{JSDM}. The concept of the two-stage precoding can be summarized as follows.
{The users are grouped into different clusters, and
each cluster corresponds to
one specific channel covariance matrix; the  downlink precoding is divided into two stages: the prebeamforming and the inner  precoding stages.
During the former stage, the prebeamforming, which only depends on the
channel covariance matrices, is utilized to eliminate the inter-cluster interference, and partitions the high dimensional massive MIMO links into several independent equivalent channels of small sizes; during the latter one, each cluster separately performs the inner precoding to eliminate the
intra cluster interferences.}
 {Moreover, {Adhikary \emph{et al.}} pointed out that
the two-stage precoding can effectively reduce the overhead of
both the downlink training and the uplink CSI feedback \cite{JSDM}.
}

 {Recently, several   works  about the JSDM have been  reported.}
In \cite{JSDM},  {the block diagonalization (BD) algorithm was proposed to
 {derive} the  prebeamforming matrix through projecting the eigenspace of channel covariance for the desired cluster onto the nullspace of the  eigenspace for all  the  other clusters.}
In \cite{JSDM_Opportunistic},
 {Nam \emph{et al.} extended the results in \cite{JSDM},
addressed some practical issues, and designed a low-cost opportunistic user selection and prebeamforming algorithm to achieve  the  optimal sum-rate.}
In \cite{JSDM_mm},  {Adhikary \emph{et al.} improved the JSDM scheme that only requires \textcolor{blue}{statistical CSI} to decrease
the computational complexity.}
In \cite{prebeamform2}, Liu and Lau  designed a phase-based  {prebeamforming}
to maximize the minimum average data rate of  {the users}.
With such method, the number of  the  radio frequency (RF) chains can be significantly reduced. In \cite{TQF}, an iterative algorithm was proposed to  {obtain the} prebeamforming
   {to maximize} the signal-to-leakage-plus-noise ratio (SLNR).  In \cite{iterative},  Chen and Lau developed a  {low-complex} online iterative algorithm to track the prebeamforming matrix.  {Sun \emph{et al.}
  considered {the users with multiple antennas, and derived the upper bound on the ergodic achievable sum-rate\cite{beam_division}.}
  Then, {the beam division multiplex access} (BDMA) was proposed for the FDD massive MIMO system, where
  {only the statistics of the CSI was} utilized
  for the optimal downlink transmission.

However, the  works  in \cite{JSDM,JSDM_Opportunistic,JSDM_mm,prebeamform2,iterative,TQF,beam_division}
only considered {the} single-cell scenario.
\textcolor{blue}{If the multi-cell scenario is examined, we will face the inter-cell interference (ICI),
which does not exist in the single-cell network and will degenerate the performance of the
cell-edge users.}
To mitigate the impact of  the  ICI,
several
coordinated transmission schemes, such as the coordinated multipoint (CoMP) transmission  \cite{comp} and the interference alignment (IA) \cite{IA_BASE,Zhao11},
have been proposed for the classical multi-cell MIMO  {networks}.
However, their  {direct} extensions to the multi-cell massive MIMO  are  not straightforward and  {not feasible}, since  the  {achieving}
of  the  global CSI  {will consume} unaffordable  wireless transmission and backhaul resources  {if} the number of  antennas is large.
 {With the two-stage precoding},
a  {novel} coordinated transmission scheme was proposed for
 {the multi-cell networks} \cite{muticell1,muticell2,muticell3}, where the ICI is eliminated through  scheduling  {user clusters}  {into the} non-overlapping  beams.  {Because}  only the second order  {statistics of the CSI} are shared among  the  cooperating BSs, the two-stage precoding based coordinated transmission scheme  {possesses} low overhead and  is  well suited for  the  ICI mitigation in  the  FDD multi-cell massive MIMO networks.

However, the  schemes in \cite{muticell1,muticell2,muticell3} simply treat all the clusters in the coordinated cells together and design prebeamforming matrices to eliminate the inter-cell and the inter-cluster interference{\color{blue},} but  do not distinguish between the cell-edge clusters and the cell-edge clusters. As a result, the following problems still  exist.
\subsubsection{  The  unfair service for the cell-edge clusters}
In  the  classical cellular networks,
the cell-edge users suffer from
very low throughput due to  {their} serious path loss.
Under the  two-stage precoding framework,
\textcolor{blue}{the rank of the  effective equivalent  channel for the cell-edge cluster is smaller than that of the cell-center cluster, and
the   unfairness will become  more serious.
We will carefully analyze this point in Section II.B.}

\textcolor{blue}{\subsubsection{The challenges during introducing CoMP to the massive MIMO system} Through coordinating and combining signals from  the multiple BSs,
	CoMP can turn the signal interference at the cell edge into the useful signal and help the operators optimize their networks.
	Hence, the cell-edge users can obtain a more consistent service experience.
	In the conventional MIMO system with a few antennas,
	it is feasible to acquire the global CSI to perform CoMP.
	Unfortunately, in  the massive MIMO system,
	the dimension of the
	channel matrices greatly increase, and acquiring the global CSI is infeasible due to its unacceptable overhead.
	As a result, the existing CoMP schemes for the
	conventional MIMO system can not be  directly adopted  for the  massive MIMO system.}
\subsubsection{ The  serious  angle of departure (AoD)  ranges overlap between different clusters}
The two-stage precoding assigns  the  non-overlapping beams for the clusters with  the  non-overlapping AoD ranges to realize orthogonal transmission. In  the  multi-cell scenario,  the overlap of the AoD ranges   will happen with greater probability
 if  all  the clusters  are considered together,
which will seriously degrade the performance of  the  two-stage precoding.

 To solve these problems,
we  will propose  an IA-SSR based transmission scheme
under the two-stage precoding framework, where different transmission schemes are applied
for  the cell-edge  and cell-center clusters.
The main ideas of  the  IA-SSR scheme are summarized as follows.

\begin{itemize}
\item  {To deal with} the unfair service  problem,
we  introduce the IA method  to enhance the transmission for the cell-edge users.
However, the achieving of the global CSI  for the  IA  will lead to unacceptable overhead in  the  massive MIMO  systems.
    Since  the two-stage precoding can sufficiently reduce
   the dimensions  of  the  effective equivalent channels,   the combination of  the IA and the two-stage precoding   makes  cooperative transmission possible for cell-edge users  with affordable overhead.
\item To address the  overlap of the AoD ranges, we put forward   an  SSR scheme, where a  low-level transmission power is allocated   to   the  cell-center users. Since the distance between  different  cell-centers areas is  long  enough to ensure  the  large path loss, the mutual interference  between  two clusters in different cell-centers can be ignored  even if they may share the same AoD range.
\item To further maximize the sum-rate, we would like to develop a power allocation policy for the proposed IA-SSR scheme. Since the  transmission power for  the  cell-center users should be limited within a lower level, the optimal power  allocation solution can be obtained by  the  golden section search algorithm with the  water-filling method in the inner loop.
\item    A low-cost training scheme is also developed to estimate   the    effective equivalent  channels and   the covariances of the equivalent noise.
\end{itemize}


The rest of this paper is organized as follows.
The system model and problem formulation  are  described in Section II. Section III illustrates  the  main  ideas  of  the  proposed IA-SSR based transmission scheme and the  optimal power allocation policy.
The pilot design and  the  channel estimation for  the  IA-SSR are presented in Section IV. \textcolor{blue}{Some practical issues in IA-SSR implementation are discussed in Section V.} The  numerical results are given in Section   VI, and the conclusions are drawn in Section VII.

Notations:  We  use lowercase (uppercase)  boldface to  denote  vector (matrix).
$(\cdot)^T$, $(\cdot)^*$, and $(\cdot)^H $ represent  the  transpose,  the
complex conjugate and  the  Hermitian transpose, respectively. $\mathbf{I}_N$
representes a $N\times N$ identity matrix.   $\mathbb E \{\cdot\}$ means  the  expectation operator. We use $\text{tr}\{\cdot\}$, $\det\{\cdot\}$ and $\text{rank}\{\cdot\}$  to  denote the  trace,  the  determinant, and  the  rank of a matrix, respectively. $[\mathbf{X}]_{ij}$ is the $(i,j)$-th entry  of  $\mathbf{X}$. $\mathbf{n} \sim \mathcal{CN}(0,\mathbf{I}_{N})$ means  that  $\mathbf{n}$ is complex circularly-symmetric Gaussian distributed with zero mean and covariance $\mathbf{I}_{N}$.

\section{System Model }

{In this section, \textcolor{blue}{we introduce the system configuration and the spatially correlated channel model in the multi-cell massive MIMO networks.}}
\subsection{{System Configuration}}
 {Consider}
the typical three-cell {network} to implement full spectrum reuse,
where each cell consists of one  BS at the geometric center position.
Each BS is equipped with
{$N_t\gg1$} antennas in the form of uniform linear array (ULA).
The corresponding BSs are separately denoted  as ${\it BS}_1$, ${\it BS}_2$, ${\it BS}_3$.
\textcolor{blue}{We divide each cell into
six $60^\circ$ sectors, which have been treated
as  an economically attractive solution to increase the system capacity in WCDMA and LTE networks~\cite{muticell1,muticell2}.
 It is assumed that the sector antennas of 60 degrees opening are used such that the  energy of each sector would not radiate out of its angle range.}
\textcolor{blue}{ Thus, as illustrated in \figurename{ \ref{fig:system}}, only the area consisted of the  three
adjacent sectors with mutual interference  are analyzed for simplicity.}
Users, each with $N_r$-antennas,
are  partitioned
into $J$ clusters, and the ones in the same   cluster are almost
co-located.  The $k$-th user in the cluster $j$
is denoted as $UE_{j,k}$,
$k=1,2,\ldots, K_{j}$, and $j=1,2,\ldots,J$, where $K_{j}$ is the number of users in the cluster $j$.

\begin{figure}[!t]
\centering
\includegraphics[width=70mm]{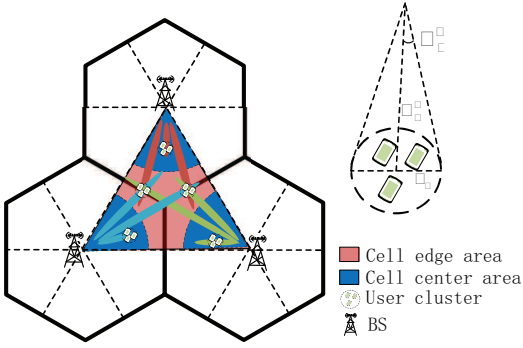}
\caption{\textcolor{blue}{The Schematic of the IA-SSR  based cooperative transmission scheme for the multi-cell  massive MIMO systems.}}
\label{fig:system}
\end{figure}
\subsection{\textcolor{blue}{Spatially Correlated Channel Model in the Massive MIMO Networks}}
The downlink massive MIMO channels from $BS_{i}$ to the user $UE_{j,k}$
can be  denoted  as the  $N_r\times N_t$ matrix
$\mathbf{H}_{j,k}^{i}$,  {which}
are assumed to be block fading.
Consider the classical  ``one-ring"
\footnote{It should be noted that the   one-ring scattering model is considered in this paper for mathematical convenience.
	The proposed transmission scheme is also applicable to the more practical scenarios with multiple-ring}
 model \cite{onering,onering2},
where the cluster $j$, located at $D_{j}^{i}$ meters away from ${BS}_{i}$,
is surrounded by a ring of scatterers with the radius $R_{j}$.
Then, for the cluster $ j $, the  angle spread (AS) at  ${BS}_{i}$ is
\begin{align}
\Delta _{j}^{i} = \arctan(R_{j}/D_{j}^{i}).\label{eq:Delta_j^i}
\end{align}

\textcolor{blue}{
	Similar to the works  \cite{iterative,prebeamform2,TQF,beam_division},
	the BSs are elevated at a very high amplitude, such that
	there is not enough local scattering around the BS antennas.
	In such cases, the  spatial correlation introduced by BS antennas should be considered.  Furthermore, it is reasonable to assume that channel matrices of different users are independent. Then,  $\mathbf{H}_{j,k}^{i}$  can be represented by
	\cite{Correlation_Models}
	\begin{align} \label{massive MIMO channel1}
	\left[\mathbf{H}_{j,k}^{i}\right]^T=
	\sqrt{\beta_{j,k}^{i} }
	{\left(\mathbf{R}_{j,k}^{i}\right)}^{1/2}
	\mathbf W_{j,k}^{i}(\mathbf{\Phi}_{j,k})^{T/2},
	\end{align}
	where $\beta_{j,k}^{i} $ is the large-scale fading coefficient; $\mathbf{R}_{j,k}^{i}$  and  $\mathbf{\Phi}_{j,k}$ are the spatial correlation  matrices at  $BS_i$  and $UE_{j,k}$, respectively;
}
{$\mathbf W_{j,k}^{i}$} is $r_{j}^{i}\times N_r$ is a random matrix, whose entries are i.i.d complex Gaussian distributed with zero mean and unit variance.
%
It is a fact that all the users in the same cluster share the same
one-ring model parameters. As a result, the spatial  correlation  matrix   satisfies  $\mathbf R _{j,k}^{i} = \mathbf R _{j}^{i}$ for all  the  users in the cluster $j$. Following the methods in\cite{TQF},
we can derive
\begin{equation}\label{correlation}
\left[\mathbf{R}_{j}^{i}\right]_{p,q}  =
\frac{1}{2\Delta _{j}^{i}} \int _{\theta_{j}^{i} -\Delta_{j}^{i} }^{\theta_{j}^{i} +\Delta_{j}^{i}}
e^{\frac{-2i\pi (p-q)  \sin(\alpha) \tau}{\lambda}} d\alpha{\color{blue},}
\end{equation}
where $\theta_{j}^{i}$ is the azimuth angle {corresponding to the} central point of scatters ring, $ \tau$ is the antenna element spacing, and $\lambda$ is the carrier wavelength.

Resorting to eigen-decomposition, we can obtain
\begin{align}\label{def-R}
\mathbf{R}_{j}^{i} = \mathbf{E}_{j}^{i}    {\bm{{\Lambda}}_{j}^{i}   }     \big(\mathbf{E}_{j}^{i} \big)^H,
\end{align}
where
{${\bm{{\Lambda}}_{j}^{i}}$} is an
$r_{j}^{i}\times r_{j}^{i}$ diagonal matrix with the
nonzero eigenvalues of $\mathbf{{R}}_{j}^{i}$ as the main diagonal elements,
$\mathbf{E}_{j}^{i}$ is the $N_t\times r_{j}^{i}$ tall unitary matrix constructed by
the eigenvectors of  $\mathbf{R}_{j}^{i} $ corresponding to the
nonzero eigenvalues, and $r_{j}^{i} $ denotes the rank of $\mathbf{R}_{j}^{i} $. {{With the Karhunen-Loeve representation, the matrix}}
$\big[\mathbf{H}_{j,k}^{i}\big]^T$ can be  re-expressed  as~\cite{JSDM}
\textcolor{blue}{
	\begin{align} \label{massive MIMO channel}
	\left[\mathbf{H}_{j,k}^{i}\right]^T=
	\sqrt{\beta_{j,k}^i} \mathbf E_{j}^i
	{\left(\bm{{\Lambda}}_{j}^{i}\right) }^{\frac{1}{2}}
	\mathbf W_{j,k}^{i}(\mathbf{\Phi}_{j,k})^{T/2}.
	\end{align}
}
{Moreover, we define the $N_rK_{j}\times N_t$ matrix
$\mathbf{H}_{j}^{i} = \big[ (\mathbf{H}_{j,1}^{i})^T, (\mathbf{H}_{j,2}^{i})^T, \cdots,(\mathbf{H}_{j,K_{j}}^{i})^T  \big] ^T$  as
the downlink channel matrix from the $BS_i$ to the cluster $j$.}

\textcolor{blue}
{Interestingly, it can be readily checked that
	$
	\left[\mathbf{R}_{j}^{i}\right]_{p,q}  = \left[\mathbf{R}_{j}^{i}\right]_{p+1,q+1}
	$,  which means that  $\mathbf{R}_j^i$ is a  Toeplitz matrix. In the massive MIMO system, as $N_t$  approaches the infinity,  $\mathbf{R}_{j}^{i} $ asymptotically tends to be a circulant matrix,} and $\mathbf{E}_{j}^{i}$ can be constructed by $r_{j}^{i}$ columns of the $N_t \times N_t$ unitary discrete Fourier transform (DFT) matrix $\mathbf F_{N_t}$ as \cite{gao_E}
\begin{align}
\!\!\mathbf{E}_{j}^{i} =\big[ \mathbf f_n : n \in \mathcal{I}_j^i \big]{\color{blue},}
\end{align}
where $\mathbf f_n$ represents  the  $n$-th column of $\mathbf F_{N_t}$, and {the index set $\mathcal{I}_j^i$  can be written as
\textcolor{blue}{
\begin{align} \label{rank}
\mathcal I_j^i =& \Big\{ n:2n/N_t-1  \in \big[\frac{\tau}{\lambda} \sin( {\theta_{j}^{i} \!+\Delta_{j}^{i} }). \notag\\
&~~~~~~~~~\frac{\tau}{\lambda}  \sin( {\theta_{j}^{i} \!-\Delta_{j}^{i} })\big], n=0,1,\cdots,N_t-1\Big\}\notag\\
=& \Big\{ n:n  \in \big[N_t\frac{\tau}{\lambda} \sin( {\theta_{j}^{i} \!+\Delta_{j}^{i} })+\frac{N_t}{2}, \notag\\
&N_t\frac{\tau}{\lambda}  \sin( {\theta_{j}^{i} \!-\Delta_{j}^{i} })+\frac{N_t}{2}\big], n=0,1,\cdots,N_t-1\Big\}.
\end{align}
Then,  {$r^i_j$ equals the cardinality of the index set $\mathcal{I}_j^i$, i.e.},
\begin{align}\label{rank_r}
r_{j}^{i} = &\left|N_t\frac{\tau}{\lambda} \sin( {\theta_{j}^{i} +\Delta_{j}^{i} })-N_t\frac{\tau}{\lambda} \sin( {\theta_{j}^{i} -\Delta_{j}^{i} })\right|\notag\\
= &2N_t\frac{\tau}{\lambda}\left| \cos( {\theta_{j}^{i}  })\right| \sin( {\Delta_{j}^{i} }) \notag\\
=&2N_t \frac{\tau}{\lambda}\left| \cos( {\theta_{j}^{i}  })\right| \sin\big( \arctan(R_{j}/D_{j}^{i})\big).
\end{align}
}
Since the AS $\Delta _{j}^{i} $ is relatively small,
$\mathbf{R}_{j}^{i} $ possesses low rank property, i.e., $r_{j}^{i}\ll N_t$.
%
%
%
\begin{figure}[!t]
\centering
\includegraphics[width=80mm]{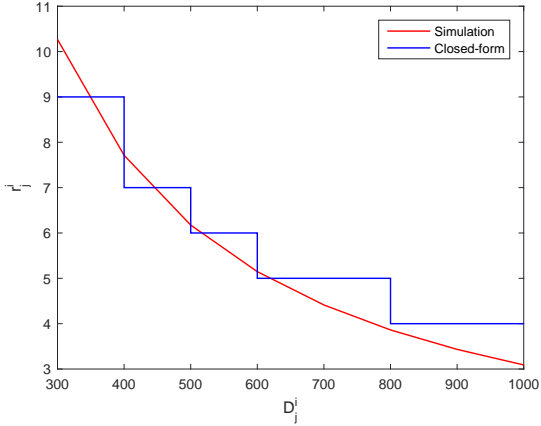}
\caption{   $r_{j}^{i}$   versus    $D_{j}^{i}$  with $ N_t =128,\theta_{j}^{i} = 0^{\circ}$, and $ R_{j} = 25m$.}
\label{fig:rank}
\end{figure}
 {With} \eqref{rank_r},   {we can present} the following  theorem.

\textcolor{blue}{
\begin{theorem} \label{th1}
In the one-ring channel model, if the parameter $R_j^i$ is fixed,
the rank $r_j^i$ decreases with the increasing of the  distance $D_j^i$.
\end {theorem}}

To better understand Theorem \ref{th1}, we present the curves of $r_{j}^{i}$ versus  $D_{j}^{i}$ in { \figurename{ \ref{fig:rank}}}, where
$ N_t =128$, $\theta_{j}^{i} = 0^{\circ}$, and $ R_{j}  = 25m$.
The red and blue curves are obtained from  the  closed-form expression in \eqref{rank_r} and
Monte Carlo numerical simulations, respectively.
\figurename{ \ref{fig:rank}}} shows that
the rank $r_{j}^{i}$ of $\mathbf R_j^i$ equals   9 for the clusters with $D_{j}^{i}=300 m$, but
reduces to 4 when $D_{j}^{i}$ increases to $900 m$,
which means that  {the  rank of the channel covariances} for the users in the cell-edge areas  is obviously smaller than that  {for the} cell-center users.

\section{Proposed IA-SSR { Based} Cooperative Transmission Scheme}
\textcolor{blue}{In this section, the proposed  transmission scheme will be presented}
under the two-stage precoding framework.}
\figurename{ \ref{fig:system}}  shows the main idea of our proposed IA-SSR  scheme.
The coverage area of each sector is divided into the cell-center area and the cell-edge area. Correspondingly, the total coverage area can be partitioned into three different cell-center areas and one big cell-edge area, which consists of the edge areas of all the three sectors and is marked in red.
Let us collect the clusters located in the cell-center area $i$ into the set $\mathcal C_i$, and gather the clusters  within the cell-edge area into the set $\mathcal E$,  {i=1,2,3.}
To fully exploit the spatial DoF  of {the two-stage precoding},
we {will} design different transmission  schemes  for {the} cell-edge  and {the} cell-center clusters.

\subsection{IA Scheme for  {the} Cell-edge Cluster}

Obviously, the cluster in $\mathcal E$ has similar distance from each BS, and can be served by any  of the three BSs. \textcolor{blue}{Hence, we let three BSs  simultaneously transmit data to each cluster in $\mathcal E$ to enhance its data rate.
For analysis simplicity, the number of the users in each  cluster $j \in \mathcal E$  is assumed to be the  same as that of BSs\footnote{The scheme can be extended to the general case,  where the number of the  users in a cluster is greater than that of BSs. A simple but effective method is to allocate orthogonal time slots for the extra users.
 For example, if there are four users numbered as 1,2,3 and 4, the users can be scheduled with equal opportunity  into different slots as  $(1,2,3),(1,2,4),(1,3,4),(2,3,4)\cdots$.}.
Moreover, the data streams from each BS is intended for one specific user in  cluster $j$.}
 It should be pointed out that the cluster in $ \mathcal C_i$ is still  served only  by   $BS_i$. Under IA-SSR, the received signal of the cluster $j$ in $\mathcal E$ can be written as
\begin{align}\label{rec_edge}
\mathbf{y}_{j\in \mathcal E}
=&\sum_{i =1 }^{3} \mathbf{H}_{j}^{i} \mathbf x^{i}+ \mathbf{n}_{j} \notag \\
=&\sum_{i=1}^{3}  \mathbf{H}_{j}^{i}  \mathbf{P}_{j}^{i} \mathbf{d}_{j}^{i}
+\sum_{i=1}^{3}\sum_{j' \in \mathcal E,j'\neq j}	\mathbf{H}_{j}^{i} \mathbf{P}_{j'}^{i} \mathbf{d}_{j'}^{i} \notag \\ &+\sum_{i=1}^{3}\sum_{j'\in \mathcal C_i}	\mathbf{H}_{j}^{i} \mathbf{P}_{j'}^{i}  \mathbf{d}_{j'}^{i}
+ \mathbf{n}_{j},
\end{align}
where the $N_t\times 1$ vector $\mathbf x^i = \sum\limits_{j \in \mathcal E \cup \mathcal C_i} \mathbf{P}_{j}^i\mathbf{d}_{j}^i$ is the transmitted signal vector from  $BS_i$, $\mathbf d_{j}^i$ is the
$S^i_j \times 1$ data vector  from the $BS_i$ to the cluster $j$, $\mathbf P_{j}^i$ is the $N_t\times S_j^i$  precoding matrix for the cluster $j$ at $BS_i$, and the $N_rK_j\times 1$ vector $\mathbf{n}_{j}\! \sim \!\mathcal{CN}(\mathbf 0,\mathbf{I}_{N_rK_j})$ is the additive complex Gaussian noise.
Clearly, the first part on right hand side (RHS) of \eqref{rec_edge} is the superposition of the desired signals from the three BSs;
the second and third parts are the inter-cluster interferences  caused by the  signals  for other clusters in $\mathcal E$
and  for clusters in $\mathcal C_1$ $\mathcal C_2$ and $\mathcal C_3$, respectively.

In this paper, we adopt the two-stage precoding framework,  {and divide the
the precoding process into two stages as}
\begin{align}
\mathbf P_{j}^i=\mathbf B_{j}^i\mathbf V_{j}^i ,
\end{align}
where the $N_t\times M_{j}^i$ prebeamforming matrix $\mathbf B_{j}^i$, related to spatial correlation matrices,  is utilized to eliminate the inter-cluster interferences;
the $M_{j}^i\times S_{j}^i$ matrix $\mathbf V_{j}^i$ denotes
the inner precoder dealing with the intra-cluster interferences,
{and depends on} $K_{j}N_r\times M^{i}_{j}$ effective equivalent  channel  matrix $\overline{\mathbf{H}}_j^{i} = {\mathbf{H}}_{j}^{i} \mathbf{B}^i_{j}$;
$M_{j}^i$ is the rank of $\overline{\mathbf{H}}_j^{i}$ seen by the inner precoder,  {and} satisfies  {the constraint} $S_{j}^i \leq M_{j}^i \leq r_{j}^i$.

The design  of the prebeamforming matrix $\mathbf B^i_j$
has been examined in \cite{JSDM,JSDM_Opportunistic,JSDM_mm,prebeamform2,iterative,TQF,beam_division}.
Without loss of generality, we adopt the DFT based prebeamforming,
\footnote{ \textcolor{blue}{In order to explain our core ideas  concisely,
	we adopt the classical DFT based prebeamforming to achieve $\mathbf B^i_j$.
Nonetheless, the other prebeamforming designing  methods can be  also  applicable for the proposed IA-SSR framework.} }
and achieve
the prebemforming matrices through concentrating the subspace $\text{span}\{ \mathbf B^i_j \}$
into the null-space of $\text{span}\{ \mathbf{\Xi}_{ j}^{i} \}$,
where $\mathbf{\Xi}_{ j}^{i}$  is constructed
by $\mathbf E_{j'}^i$ of all but the cluster $j$ in the system as
\begin{equation}\label{Xi}
{\mathbf{\Xi}_{j}^{i} } =\Bigg[  \mathbf f_n : n \in \bigcup_{j'=1,j'\neq j}^J \mathcal I_{j'}^i\Bigg].
\end{equation}
With the orthogonality between any two columns of the DFT matrix, ${\mathbf{B}_{j}^{i} }$ for cluster $j$ in $\mathcal E$   can be derived as
\begin{equation}\label{B_edge}
{\mathbf{B}_{j \in \mathcal E}^{i} } =\Bigg[  \mathbf f_n : n \in\Bigg( \mathcal I_{j}^i - \bigcup_{j'=1, j'\neq j}^J \mathcal I_{j'}^i \Bigg)\Bigg],
\end{equation}
where the set $\mathcal A-\mathcal B$ contains all the elements that are in the  set $\mathcal A$ but not in  the  set $\mathcal B$, i.e., $\mathcal A - \mathcal B = \{x : x \in \mathcal A \text{ and } x \notin \mathcal B\}$.
\textcolor{blue}{From the computation of $\mathbf B_j^i$, we can obtain that $M_j^i$  equals the  amount of the corresponding columns  in  $\mathbf E_j^i$  linearly independent on the ones   of  $\mathbf \Xi_j^i$
 as}
\begin{equation}\label{M_bound_without_ssr}
M^{i}_{j} = \text{rank}\{\mathbf B_j^i\}
=\Bigg| \mathcal I_{j}^i - \bigcup_{j'=1, j'\neq j}^J \mathcal I_{j'}^i \Bigg|,
\end{equation}
where $|\mathcal A|$  denotes  the number of elements in  the set $\mathcal A$. The resultant prebeamforming matrices $\mathbf B^i_j$  satisfy  the following constraint as
\begin{align}
&\left\lbrace \mathbf{E}_{j'}^{i} \right\rbrace ^H {\mathbf{B}^i_{j} } = 0 , \forall   j'\neq j,\label{BD conditon1}
\end{align}
which means that the transmitted signal to  the cluster $j \in \mathcal E$ will not cause interference to the other clusters.\footnote{
Meanwhile, the $ \mathbf B^i_j $ for cluster in cell-center areas is also designed to avoid interference to cell-edge clusters, which will be given in next subsection.}
Then the inter-cluster interference terms in (\ref{rec_edge}) are eliminated, and the received signals can be simplified as
\begin{align}\label{pb_rcv_edge}
\mathbf{y}_{j\in \mathcal E}
=&\sum_{i=1}^{3}  \mathbf{\overline{H}}_{j}^{i} \mathbf{V}_{ j}^{i} \mathbf{d}_{ j}^{i}
+\mathbf{n}_{ j}.
\end{align}

The next task is to recover the intended data from  the  superimposed
signals  of the three BSs.
We can separately present the received signal of each user as
\begin{align}
\mathbf{y}_{j,1}
=&  \overline{\mathbf{H}}_{j,1}^{1}  \mathbf{V}_{j}^{1} \mathbf{d}_{j}^{1}
+\sum_{i=2,3}  \overline{\mathbf{H}}_{j,1}^{i}  \mathbf{V}_{j}^{i} \mathbf{d}_{j}^{i}
+\mathbf{n}_{j,1},\label{y_j1}\\
\mathbf{y}_{j,2}
=&  \overline{\mathbf{H}}_{j,2}^{2}  \mathbf{V}_{j}^{2} \mathbf{d}_{j}^{2}
+\sum_{i=1,3}  \overline{\mathbf{H}}_{j,2}^{i}  \mathbf{V}_{j}^{i} \mathbf{d}_{j}^{i}
+\mathbf{n}_{j,2},\label{y_j2}\\
\mathbf{y}_{j,3}
=&  \overline{\mathbf{H}}_{j,3}^{3}  \mathbf{V}_{j}^{3} \mathbf{d}_{j}^{3}
+\sum_{i=1,2}  \overline{\mathbf{H}}_{j,3}^{i}  \mathbf{V}_{j}^{i} \mathbf{d}_{j}^{i}
+\mathbf{n}_{j,3}\label{y_j3},
\end{align}
where  $\mathbf{y}_{j,i}$ is
the received signal of $U\!E_{j\in \mathcal  E,i}$ and contains the elements in  $\mathbf{y}_{j \in \mathcal E}$ corresponding
to   $U\!E_{j\in \mathcal  E,i}$. Obviously, the first
terms on the right hand side of \eqref{y_j1}-\eqref{y_j3} are  the
desired signals  for each user, while the second terms represents the  interference.  With prebeamforming and cooperative transmission,
the equivalent channel  links  from $BS_1$, $BS_2$ and $BS_3$ to the cluster $j\in\mathcal E$  become
one three-BS three-user MIMO interference channel\cite{MIMOX},
where IA can be used to fully exploit the
spatial freedoms.

\textcolor{blue}{
Under the  IA framework,
the precoding matrix $\mathbf{V}_{j}^{i}$ can be carefully chosen to compact the interference
into one reduced-dimensional subspace  at the user side,
but to keep the desired  signals in another  subspace.
Thus, $\mathbf{V}_{j}^{i}$  should  satisfy the  following equation  set \cite{IA_BASE}:
\begin{align}
\overline{\mathbf{H}}_{j,1}^{3}  \mathbf{V}_{j}^{3}  =&  \overline{\mathbf{H}}_{j,1}^{2}  \mathbf{V}_{j}^{2}, \label{v1} \\
\overline{\mathbf{H}}_{j,2}^{3}  \mathbf{V}_{j}^{3}  =&  \overline{\mathbf{H}}_{j,2}^{1}  \mathbf{V}_{j}^{1}, \label{v2}\\
\overline{\mathbf{H}}_{j,3}^{1}  \mathbf{V}_{j}^{2}  =&  \overline{\mathbf{H}}_{j,3}^{1}  \mathbf{V}_{j}^{1}.\label{v3}
\end{align}
Correspondingly, the columns of the decoding matrix  $\mathbf{U}_{j,i}$ should be orthogonal to the subspace of the interference.  Then, each user can derive $\mathbf{U}_{j,i}$ as
\begin{align}
\mathbf{U}_{j,i} = \text{NULL}\left\lbrace \sum_{k\neq i} \overline{\mathbf{H}}_{j,i}^{k}  \mathbf{V}_{j}^{k}   \right\rbrace \label{cal_U},
\end{align}
where NULL$\{\mathbf X\}$ represents the nullspace of $\mathbf X$.
}

To ensure that IA is feasible, the following conditions should be satisfied \cite{IA_CON}{\color{blue}.}
\begin{align}
\text{rank}\big\{(\mathbf{U}_{j,i})^H \overline{\mathbf{H}}_{j,i}^{i}  \mathbf{V}_{j}^{i} \big\} = S_{j}^{i}, \label{IA_CON1}\\
(\mathbf{U}_{j,i})^H\overline{\mathbf{H}}_{j,i}^{i'}  \mathbf{V}_{j}^{i'} = \mathbf{0},\forall i \neq i'.\label{IA_CON2}
\end{align}
After the IA operation,
the interference is completely eliminated, and the received
signal at the user $ < \!\!j\!\!\in\!\mathcal E,i\!>$ can be expressed as
\begin{align}\label{eff-edge}
\underline{\mathbf{y}}_{j\in\mathcal E,i}
=\underline{\mathbf{H}}_{j,i}^{i} \mathbf{d}_{j}^{i}
+\underline{\mathbf{n}}_{j,i}{\color{blue},}
\end{align}
where $\underline{\mathbf{y}}_{j,i} = (\mathbf{U}_{j,i})^H  {\mathbf{y}}_{j,i}$,
$\underline{\mathbf{H}}_{j,i}^{i} =(\mathbf{U}_{j,i})^H \overline{ \mathbf{H}}_{j,i}^{i}  \mathbf{V}_{j}^{i}$
is the $S_{j}^i \times S_{j}^i$ full rank equivalent channel from $BS_i$ to the user $ < \!\!j\!\!\in\!\mathcal E,i\!>$,
and the distribution characteristics of
the equivalent noise $\underline{\mathbf{n}}_{j,i} = ({\mathbf{U}}_{j,i})^H {\mathbf{n}}_{j,i}$  {are the
same with that of ${\mathbf{n}}_{j,i}$}.

\begin{remark}
Under the IA based cooperative transmission, the transmitted signals from  the three
$BS$s  to  the  cluster $j'(j'\neq j)$ are designed to avoid interference to  the  cluster $j \in \mathcal E$.
Thus, it can be concluded that if  the  $BS_1$ is transmitting data to cluster $j $, $BS_2$ and $BS_3$ can also  transmit data to  the  cluster $j$,  and cause   no impact on the   other clusters they are serving.
Therefore, the IA based cooperative transmission for  the
cell-edge clusters can fully exploit spatial DoF  of the  two-stage precoding in   a  multi-cell system.
\end{remark}

\begin{remark}
IA is a promising interference management technology for a multi-cell cellular system. However, BS requires  the global CSI, which will lead to unaffordable signaling overhead in  the  massive MIMO system due to its  high dimensional channels. Fortunately,
 the  two-stage precoding can significantly reduce the equivalent channel  dimensions, which makes it possible to perform IA on  the  low dimensional equivalent channels $\overline{\mathbf{H}}_j^{i} $.
 Then, IA improves the data rate for cell-edge users under two-stage  precoding.  It can be concluded that the incorporation  between IA and  the  two-stage precoding is a potential interference management technology  with an affordable signal overhead  over the  multi-cell massive MIMO systems.
\end{remark}

\textcolor{blue}{
\begin{remark}
In contrast with existing cooperation methods \cite{comp1,comp2,comp}, which  share
both CSI and user data  streams among different BSs,
the proposed scheme only shares CSI  and can
reduce the  amount of  signaling overhead  along
the backhaul links between the BSs.
\end{remark}
}

To meet conditions \eqref{IA_CON1} and \eqref{IA_CON2},
the number of  the  data streams  should satisfy \cite{IA_CON}
\begin{align}
&S^i_j \leq \min(M^i_j , N_r),\label{IA_constrains1}\\
&\sum_{i=1}^{3} S^i_j(M^i_j+ N_r-2S^i_j) \geq S^1_j S^2_j + S^2_j S^3_j + S^1_j S^3_j,\label{IA_constrains2}\\
& S^i_j   +  S^{i'}_j \leq\notag\\
& \min  \big( M^i_j  + M^{i'}_j , 2N_r,
\max(M^i_j , N_r) ,\max(M^{i'}_j, N_r) \big).\label{IA_constrains3}
\end{align}
Then, the optimal spatial DoF can be obtained by solving the following optimization problem:
\begin{align}
(\text P1)~~~~~~~~&\max_{S^1_j,S^2_j,S^3_j} S^1_j + S^2_j +  S^3_j\notag \\
&~~{\text{s.t.}}\text{$~~~$\eqref{IA_constrains1}, \eqref{IA_constrains2}, \eqref{IA_constrains3}}, \notag
\end{align}
whose solution can be found  the  exhaustive search. The \textcolor{blue}{corresponding
search complexity is $\prod\limits_{i=1}^{3}\big[\min(M^i_j,N_r)\big]$ and affordable.}

\subsection{SSR Scheme for  the  Cell-center Cluster}
In  the  IA-SSR, the received signal of the cluster $j$ in $\mathcal C_i$ can be written as
\begin{align}\label{rec_center}
&{\mathbf{y}_{j \in\mathcal C_i}}
=\mathbf{H}_{j}^{i} \mathbf{B}_{j}^{i} \mathbf{V}_{j}^{i} \mathbf{d}_{j}^{i}
+\sum_{j'\in\mathcal C_i,j'\neq j}	\mathbf{H}_{j}^{i} \mathbf{B}_{j'}^{i} \mathbf{V}_{j'}^{i}\mathbf{d}_{j'}^{i}\notag\\
&+\sum_{i'=1}^{3}\sum_{j'\in \mathcal E}\!\!	\mathbf{H}_{j}^{i'} \mathbf{B}_{j'}^{i'} \mathbf{V}_{j'}^{i'} \mathbf{d}_{j'}^{i'}
+\!\!\!\!\sum_{i'\!=1,i'\!\neq i}^{3}  \sum_{j'\in \mathcal C_{i'}}\!\!	\mathbf{H}_{j}^{i'} \mathbf{B}_{j'}^{i'} \mathbf{V}_{j'}^{i'} \mathbf{d}_{j'}^{i'} \!\!+ \!\mathbf{n}_{j} ,
\end{align}
where the first part on  the  RHS of the above equation is the desired signal;
the second and third parts are the interferences caused by the streams from $BS_i$ to the other clusters
in $\mathcal C_i$ and from $BS_1$, $BS_2$ and $BS_3$ to the clusters in $\mathcal E$, respectively; the fourth part is the interference caused by streams for clusters in the  other two cell-center areas.

{The main operations of SSR can be presented as follows.
The $BS_i$ loads one high-level power on the   streams from the $BS_i$ to the clusters in $\mathcal E$,
but assigns one low-level power for the streams to the clusters in $\mathcal C_i$.
Since the distance between  two  cell-center areas is long  enough to ensure  the  large path loss,
the    mutual-interference  between two clusters in different cell-centers is low enough to be treated as noise.}

With SSR, the  data streams to one specific cluster in $\mathcal C_i$
should be optimized to avoid interference with the clusters in both
$\mathcal C_i$ and  $\mathcal E$ but not in $\mathcal C_{i^\prime}$, $i\neq i^\prime$.
Then, {for DFT based prebeamforming}, the subspace $\text{span}\{\mathbf B_{j\in\mathcal C_i}^i\}$
should be orthogonal with the subspace span$\{\mathbf \Psi_j^i\}$ other than $\text{span}\{ \mathbf{\Xi}_{ j}^{i} \}$ in \eqref{Xi}, where $\mathbf{\Psi}_{ j}^{i}$  is
constructed
by $\mathbf E_{j'}^i$ of  {all clusters but $j$ in $\mathcal C_i$ and those in $\mathcal E$ as}
\begin{equation}\label{Psi}
{\mathbf{\Psi}_{j}^{i} } =\Bigg[  \mathbf f_n : n \in \bigcup_{j'\in \mathcal C_i \cup \mathcal E,j'\neq j } \mathcal I_{j'}^i\Bigg] .
\end{equation}
Then, we can derive the prebeamforming matrix for the cluster $j$ in $\mathcal C_i$ as
\begin{equation}\label{B_center}
{\mathbf{B}_{j \in \mathcal C_i}^{i} } =\Bigg[  \mathbf f_n : n \in\Bigg( \mathcal I_{j}^i - \bigcup_{j'\in \mathcal C_i \cup \mathcal E,j'\neq j } \mathcal I_{j'}^i \Bigg)\Bigg] .
\end{equation}


After prebeamforming, the received signal in \eqref{rec_center} can be simplified as
\begin{align}\label{pb_rcv_center}
{\mathbf{y}_{j\in\mathcal C_i}}
=&\mathbf{\overline{H}}_{j}^{i} \mathbf{V}_{j}^{i} \mathbf{d}_{j}^{i}
+\underline{\mathbf{n}}_{j},
\end{align}
where \begin{align}\label{equivalant_noise}
\underline{\mathbf{n}}_{j} = \sum_{i'\!=1,i'\!\neq i}^{3}  \sum_{j'\in \mathcal C_{i'}}\!\!	\mathbf{H}_{j}^{i'} \mathbf{B}_{j'}^{i'} \mathbf{V}_{j'}^{i'} \mathbf{d}_{j'}^{i'} \!\!+ \!\mathbf{n}_{j}{\color{blue},}
\end{align}
denotes the equivalent noise, and   its conditional covariance matrix
on given $\{\mathbf{H}_{j}^{i'} \mathbf{B}_{j'}^{i'}\}$ is denoted by $\mathbf{K}_{j\in \mathcal C_i}$.

For given $\mathbf{B}_{j}^{i}$, $\mathbf{\overline{H}}_{j}^{i}$ represents the traditional multiuser MIMO  broadcast channel~\cite{BC},
and the ZF inner precoder can be utilized to deal with intra-cluster interference~\cite{zf}.
The detailed ZF inner precoder can be written as
\begin{align}
\mathbf{V}_{j}^{i} = \zeta_{ j}^{i}\mathbf{{Z}}_{j}^{i}\label{V_cellcenter},
\end{align}
where \textcolor[rgb]{0.00,0.00,1.00}{$\mathbf{{Z}}_{j}^{i}=\big(\mathbf{\overline{H}}_{j}^{i }\big)^H\left( \mathbf{\overline{H}}_{j}^{i} \big( \mathbf{\overline{H}}_{j}^{i}\big) ^H\right) ^{-1} $, and $\zeta_{j}^{i}=\sqrt{\frac{S_{j}^{i}}{\text{tr}\left( \mathbf{{Z}}_{j}^{i}\left( \mathbf{{Z}}_{j}^{i}\right) ^H\right)}}$ is a normalization factor to constrain the power gain of  inner precoder.}

 After  the  ZF inner precoding, the received signal of  the  cell-center  cluster in \eqref{pb_rcv_center} can be rewritten as
\begin{align}\label{eff-cent}
{\mathbf{y}_{j\in\mathbb C_i}}
=&\mathbf{\underline{H}}_{j}^{i} \mathbf{d}_{j}^{i}
+\underline{\mathbf{n}}_{j} ,
\end{align}
where $\underline{\mathbf{H}}_{j}^{i} = \overline{ \mathbf{H}}_{j}^{i}  \mathbf{V}_{j}^{i} = \zeta_{j}^{i} \mathbf{I}_{S_{j}^{i}}$ is  the  $S_{j}^{i} \times S_{j}^{i}$ equivalent channel
 for the cluster $j$ in the $\mathcal C_i$.

Under SSR, $M_j^i$  equals the number of columns in $\mathbf E_j^i$ that  are linearly independent  on the columns of  $\mathbf \Psi_j^i$.
The number of  the
data streams from $BS_i$ to  the  cluster $j \in \mathcal C_i$ is given by
\begin{align}\label{S_bound}
S^{i}_{j\in \mathcal C_i} &=M^{i}_{j\in \mathcal C_i} = \text{rank}\{\mathbf B_j^i\}=\Bigg|  \mathcal I_{j}^i - \bigcup_{j'\in \mathcal C_i \cup \mathcal E,j'\neq j } \mathcal I_{j'}^i\Bigg|.
\end{align}
Comparing with  the scenario  without SSR in \eqref{M_bound_without_ssr},
the number of data streams for  the  cell-center clusters may be significantly improved, which can be  explained as follows.
When SSR is not adopted, if  the AoD range  of  the  cluster $j \in \mathcal C_i$ overlaps with that of the cluster $j' \in \mathcal C_{i'}$  at  $BS_i$, $BS_i$  should not  transmit data streams for  the  cluster $j \in \mathcal C_i$ on the  overlapping beams in order to avoid interference. {Nonetheless},   {if we utilize} SSR,  $BS_i$ can transmit data streams to  {the} cluster $j  \in \mathcal C_{i }$,  {while}
$BS_{i'}$ can  {simultaneously transmit data streams to
	$j'  \in \mathcal C_{i' }$}.
\begin{remark}
	With  the  SSR scheme,  the AoD range overlapping  is mitigated since the mutual-interference between two clusters in different cell-centers is negligible.
	 Thus, we do not need to treat all  the  clusters in the
	whole system as one big group anymore. However, the mutual-interference
	between cell-center  clusters and cell-edge  clusters  sill exists.
	Therefore, for the cluster $j\in \mathcal E $,
	its prebeamforming should still be well
	designed to avoid interference with all the other clusters in the system.
\end{remark}
\subsection{Power Allocation for IA-SSR}
In this section, we will develop a power allocation policy for the proposed IA-SSR  scheme to
maximize the sum-capacity. Thanks to the two-stage precoding scheme,
the channel links for the whole system can be
decomposed into several independent equivalent channel links
with reduced dimensions, as shown in  \eqref{eff-edge} and \eqref{eff-cent}.
 Let us assume  that BSs have  knowledge of the effective equivalent  channels  $\{ \underline{\mathbf{H}}_{j}^{i}\}$ and  the  equivalent noise covariance  matrices $\{ \mathbf{K}_{ j \in \mathcal C_i}\}$.\footnote{We will provide a method to estimate $\{ \underline{\mathbf{H}}_{j}^{i}\}$ and $\{ \mathbf{K}_{ j \in \mathcal C_i}\}$  in  the  next subsection.}
Then, the achievable capacity of  the link  from $ BS_i$ to    the  cluster $j$  in $\mathcal E$
can be expressed as  \cite{tse2005fundamentals}
\begin{align}\label{c_sum_edge}
\mathbb C_{j \in \mathcal E,i}^{i} \left( \underline{\mathbf{H}}_{j,i}^{i}\right)
&=\log\det\Big\{\mathbf{I}_{S_j^i}  + \left( \underline{\mathbf{H}}_{j,i}^{i}\right)  \mathbf{Q}_j^i
\left( \underline{\mathbf{H}}_{j,i}^{i}\right)^H\Big\}\notag\\
&=\sum_{s=1}^{S_{j}^i}\log \left\lbrace  1 + \lambda_{j,s}^{i}  p_{j,s}^{i}   \right\rbrace,
\end{align}
where $\lambda_{j,s}^{i}$ is the $s$-th eigenvalue of   $\underline{\mathbf{H}}_{j,i}^{i} (\underline{\mathbf{H}}_{j,i}^{i})^H$, $\mathbf{Q}_j^i = \mathbb{E}\left\lbrace \mathbf{d}_{j}^{i} ( \mathbf{d}_{j}^{i}) ^H  \right\rbrace${\color{blue},}
and  {its $s$-th eigenvalue} $p_{j,s}^{i} $ represents the {transmitting power} for the $s$-th data stream from $BS_i$ to  the  cluster $j$.

Similarly, the achievable capacity of  {the link}
from $BS_i$ to   the  cluster  $j $ in $\mathcal C_i$ can be written as
\begin{align}\label{c_sum_cent}
{\mathbb C_{j \in \mathcal C_i}^i}\left( \underline{\mathbf{H}}_{j}^{i}\right)
&=\log\det\Big\{\mathbf{I}  + \mathbf{K}_{ j}^{-1} \left( \underline{\mathbf{H}}_{j}^{i}\right)  \mathbf{Q}_j^i
\left( \underline{\mathbf{H}}_{j}^{i}\right)^H\Big\}\notag\\
&= \sum_{s=1}^{S_{j}^i}\log \left\{  1 + k_{j,s}^{-1} \left(\zeta_{j}^{i}\right) ^{2} p_{j,s}^{i}   \right\},
\end{align}
where $k_{j,s} $ is $s$-th eigenvalue of $\mathbf{K}_j$.
In IA-SSR,  $BS_i$  serves  all  the  clusters in $\mathcal C_i$ and  the  $i$-th user of each  cluster in $\mathcal E$. Then, the achievable sum-capacity of the {whole} system can be listed as
\begin{align}\label{c_sum}
\mathbb C_{sum}=
\sum_{i=1}^{3}\Bigg(
\sum_{j \in \mathcal C_i}\mathbb C_{j}^i
+\sum_{j \in \mathcal E}\mathbb C_{j,i}^i\Bigg) .
\end{align}

As mentioned in  the  above subsection,
we load one {low-level} power  $p_{cent}$ on each data stream for
the cell-center clusters to avoid the mutual-interference between different cell-centers, i.e.{\color{blue},} $p_{j \in \mathcal C_i ,s}^i = p_{cent}$.
To maximize the sum-capacity of the network,  the optimization  problem {with respect to} both $p_{cent}$ and
${p}_{j\in \mathcal{E} ,s}^{i}$ can be formulated as
\begin{align}\label{max_c_sum}
(\text P2)& \max_{p_{cent},~{p}_{j\in \mathcal{E} ,s}^{i}}\mathbb C_{sum}\\\notag
\text{s.t.}~~~& \sum_{i=1}^{3}\Big(\sum_{j\in \mathcal C_i} \sum_{s=1}^{S_{j}^i} p_{cent} + \sum_{j\in \mathcal E} \sum_{s=1}^{S_{j}^i} p_{j,s}^i \Big) \leq P_{total},\\\notag
&p_{cent} \geq 0 ,\\\notag
&p_{j\in \mathcal{E},s}^i \geq 0,
\end{align}
 where $P_{total}$ is the total power.
 Unfortunately,  since the equivalent noise of the cell-center cluster is dependent on $p_{cent}$, the classical water-filling algorithm cannot be used to solve the above problem. Nonetheless, for  the  given $p_{cent}$, the problem (P2) can be  simplified as
\begin{align}\label{max_c_sum2}
(\text P3) ~~~~~~~~ \max_{{p}_{j\in \mathcal{E} ,s}^{i}}\mathbb C_{sum,\mathcal{E}} &= \sum_{i=1}^{3}
\sum_{j \in \mathcal E}\mathbb C_{j,i}^i\\\notag
{\text{s.t.}}~~~ \sum_{i=1}^{3} \sum_{j\in \mathcal E} \sum_{s=1}^{S_{j}^i} p_{j,s}^i   &\leq P_{total} - \sum_{i=1}^{3} \sum_{j\in \mathcal C_i} \sum_{s=1}^{S_{j}^i} p_{cent} ,\\\notag
p_{j\in \mathcal{E} ,s}^i &\geq 0.
\end{align}
With the Karush-Kuhn-Tucker condition, we can achieve the solution for (\text P3) as
\begin{align}
p_{j \in \mathcal E,s}^{i}  &= \max{ \left\lbrace 0, \frac{1}{\mu} - \frac{1}{\lambda_{j,s}^{i}}\right\rbrace }\label{p_E},
\end{align}
where $\mu$ is the Lagrange multiplier factor and satisfies
\begin{align}\label{mu_condition}
&\sum_{i=1}^{3} \sum_{j\in \mathcal E} \sum_{s=1}^{S_{j}^i}   \max{ \left\lbrace  0,  \frac{1}{\mu} - \frac{1}{\lambda_{j,s}^{i}}\right\rbrace }\notag\\
 &  ~~~~~~~~~~~~~~~~=  P_{total}  -  \sum_{i=1}^{3}  \sum_{j\in \mathcal C_i} \sum_{s=1}^{S_{j}^i} p_{cent}.
\end{align}
According to (42) and (43), the water-filling algorithm can be adopted to perform the power allocation  {for the cell-edge clusters} with given $p_{cent}$.

Obviously, the sum-capacity $\sum\limits_{i=1}^{3} \sum\limits_{j \in \mathcal E}\mathbb C_{j,i}^i$
of  the  cell-edge clusters  always decreases with the increase of $p_{cent}$. Moreover,
when $p_{cent}$ lies in { a} small  regime,   the sum-capacity of  the cell-center cluster  $   \sum\limits_{i=1}^{3}
\sum_{j \in \mathcal C_i}\mathbb C_{j,i}^i$ is proportional to $p_{cent}$. However, when $p_{cent}$ lies in { a} large regime, the  interference between different cell-centers will be introduced,  and the sum-capacity of  the
cell-center cluster will be  limited by the interference. Therefore, with the increase of  $p_{cent}$, $\mathbb C_{sum}$ first increases and then decreases. With this property, we \textcolor{blue}{will utilize the golden section method    to solve the power allocation problem (P2)}, which is described in Algorithm \ref{alg:1} \cite{0.618}.
\begin{algorithm}[!t]
\caption{Power Allocation Algorithm  }
\label{alg:1}
\begin{algorithmic}[1]
\STATE {Initialize  maximum tolerance $\varepsilon$, set $p_{l}=0, p_{r}=\frac{P_{total}}{\sum_{i=1}^{3}\sum_{j \in \mathcal C_i} S_{j}^i}$}.
\REPEAT
\STATE
$p_{m1} = p_{l}+ 0.382| p_{r}-p_{l} |$;\\
solve problem (P3) for a given $p_{cent} = p_{m1}$
and	obtain power allocation policies $\{{p}_{j\in \mathcal{E} ,s}^{i}\}_1$;
calculate ${\mathbb C_{sum}}_1$ through \eqref{c_sum}.
\STATE
$p_{m2} = p_{l}+ 0.618| p_{r}-p_{l} |$;\\
solve problem (P3) for a given $p_{cent} = p_{m2}$
and	obtain power allocation policies $\{{p}_{j\in \mathcal{E} ,s}^{i}\}_{2}$;
calculate ${\mathbb C_{sum}}_{2}$ through \eqref{c_sum}.
\IF{${\mathbb C_{sum}}_{1}> {\mathbb C_{sum}}_{2}$}
\STATE  $p_{l}=p_{m1}$; $p_{cent} = p_{m1};\{p_{j\in \mathcal E,s}^{i}\} = \{{p}_{j\in \mathcal{E} ,s}^{i}\}_1$.
\ELSE
\STATE  $p_{r}=p_{m2}$; $p_{cent} = p_{m2};\{p_{j\in \mathcal E,s}^{i}\} = \{{p}_{j\in \mathcal{E} ,s}^{i}\}_2$.
\ENDIF
\UNTIL{$|p_{r} - p_{l}|<\varepsilon$}
\RETURN{ $p_{cent},\{p_{j\in \mathcal E,s}^{i}\}$}
\end{algorithmic}
\end{algorithm}

\section{The  Channel Estimation   for IA-SSR}

In order to design the prebeamforming matrices, BSs need to acquire the channel eigenspaces  $ \mathbf{E}_{j}^{i}  $ of each cluster.  {From \eqref{def-R}, we know}  that the spatial correlation matrix  of a cluster is determined by its AoD and AS, which change slowly with respect to the channel coherent time {\cite{dai}}.  Moreover, the uplink-downlink reciprocity exists for spatial correlation matrix even in the FDD system \cite{gao_E2}.
Hence,  $\{\mathbf{E}_{j}^{i} \} $ can be tracked from  the  uplink training  with low overhead  \cite{gao_E3}. Here, we assume that $\{\mathbf{E}_{j}^{i} \} $  is available to BSs.

 In the  FDD system, unlike $\{\mathbf{E}_{j}^{i} \} $,
the effective equivalent  channel $\{\overline {\mathbf{H}}_{j}^{i} \} $   should be obtained through  the  downlink training and  the
user feedback within each channel coherent block, which accounts for the major part of the total signaling overhead in IA-SSR.
In order to estimate  $\overline {\mathbf{H}}_{j }^{i}$ and $\mathbf{K}_{j\in \mathcal C_i}$  in a  { low-overhead} way, we propose a training scheme  to reuse
the training matrices within each BS,  where  the linear least squares (LS)  estimator  is adopted. In our scheme, $ BS_i$ transmits the training matrices $\mathbf{ T_e}_{j}^{i} $ of size
$M_{j}^{i} \times {T_e}_j$ and $\mathbf {T_c}_{i}^{j}  $ of size $M_{j}^{i} \times {T_c}_j$, respectively.

\subsection{
 \textcolor{blue}{The Estimation of the Effective Equivalent  Channels $\{\overline {\mathbf{H}}_{j \in \mathcal{E}}^{i} \}$, $\{\overline {\mathbf{H}}_{j \in \mathcal C_i}^{i} \}$  }}
\textcolor{blue}{It can be found that $\overline{\mathbf{H}}_j^{i} $  {possesses} a much smaller number of unknown  parameters  than the original channel matrix $\mathbf{H}_{j}^{i}$.
As a result,  the estimation of  $\overline{\mathbf{H}}_j^{i} $ consumes less channel resource than  estimation of  $\mathbf{H}_{j}^{i}$.}
We first consider the estimation of $\mathbf{\overline{H}}_{j \in \mathcal{E}}^{i}$,
 {$i=1,2,3$}.
Recalling the  effective equivalent  channel model in \eqref{pb_rcv_edge}, the received training signal of the cluster $j$ in $\mathcal E$ is given by
\begin{align}\label{tr_edge}
\mathbf{Y}_{j\in \mathcal E}
=&\sum_{i=1}^{3}  \mathbf{\overline{H}}_{j}^{i} \mathbf{T_e}_{ j}^{i}
+\mathbf{N}_{ j}{\color{blue},}
\end{align}
where   $\mathbf{N}_{j} = [{\mathbf{n}_{j}}_1,{\mathbf{n}_{j}}_2,\cdots,{\mathbf{n}_{j}}_{{T_c}_j}]$ contains the noise vectors of   $ {T_c}_j$  time slots.
 {From the LS theory, to implement the optimal estimation of }
 $\{\overline{ \mathbf{H}}_{j\in \mathcal E}^{i} \}(i=1,2,3)$,
 the training matrices from the three BSs to a specific  cluster $j$ in $\mathcal E$  should be orthogonal with each other,  {which means that}
\begin{align}\label{edge_tr_condition}
\mathbf{T_e}_{ j}^{i}( \mathbf{T_e}_{ j}^{i'})^H &= \mathbf 0, i\neq i'.
\end{align}

Since  the channel links from one specific BS to  its served clusters are independent,  the training matrices can be reused among different clusters.  With this consideration,  the minimal value  of ${T_e}_{j}$ satisfying \eqref{edge_tr_condition}  is
\begin{align}
{T_e}_{j} =\sum\limits_{i=1}^{3} M_{j}^{i}.
\end{align}

Then, we  focus on the estimation of
$\mathbf{\overline{H}}_{j\in \mathcal C_i}^{i}$.
Similarly, we can obtain the received training signal of the cluster $j$ in $\mathcal C_i$  as
\begin{align}
{\mathbf{Y}_{j\in\mathcal C_i}}
=&\mathbf{\overline{H}}_{j}^{i} \mathbf{T_c}_{j}^{i}
+
\sum_{i'\!=1,i'\!\neq i}^{3}  \sum_{j'\in \mathcal C_{i'}}\!\!	\mathbf{H}_{j}^{i'} \mathbf{B}_{j'}^{i'} \mathbf{T_c}_{j'}^{i'}  \!\!+ \!\mathbf{N}_{j}.
\end{align}
To estimate $\overline{\mathbf{H}}_{j\in\mathcal C_i}^{i}$ in absence of interference, the training matrices for the clusters  in the cell-center areas should satisfy
\begin{align}\label{cent_tr_condition}
\mathbf{T_c}_{ j}^{i}( \mathbf{T_c}_{ j'}^{i'})^H &= \mathbf 0,j\in \mathcal C_i,j'\in \mathcal C_{i'}, i\neq i'.
\end{align}

Furthermore, the covariance matrix $\mathbf{K}_{j\in \mathcal C_i}$ of  the
equivalent noise in \eqref{equivalant_noise}  can be  achieved through  estimating each interference channel $\mathbf{H}_{j}^{i'} \mathbf{B}_{j'}^{i'}$,
which will need large training matrices.  To save training resources, we directly estimate the sum  of interference channels instead  of the  respective channels. Then, the  training  resources can be reused within each BS, and only the orthogonality  among
training matrices from different BSs is required to satisfy \eqref{cent_tr_condition}.  Therefore, the minimal  dimension ${T_c}_j$  can be listed as
\begin{align}
{T_c}_j = \sum_{i=1}^{3} \bar M^i,
\end{align}
where  {$ \bar M^i = \max\limits_{j\in\mathcal C_i} M_j^i $}. To make the proposed scheme more specific, we would like to  show an example of  the designed training matrices and the  whole  procedures of   channel estimation.
{Here, we construct the training matrices  for the cluster $j$ in $\mathcal E$ from the ${T_e}_j \times {T_e}_j$ DFT matrix $\mathbf  F_{e}$  as}
\begin{align}
&\mathbf{T_e}_j^1 =  \big[\mathbf  F_{e}  \big]_{1:M_j^1,} \label{train_edge_1}\\
&\mathbf{T_e}_j^2 =  \big[\mathbf  F_{e}  \big]_{M_j^1+1:(M_j^1+M_j^2),}\\
&\mathbf{T_e}_j^3 =  \big[\mathbf  F_{e}  \big]_{M_j^1+M_j^2+1:(M_j^1+M_j^2+M_j^3),}
\end{align}
where $[\mathbf A]_{m:n}$ {denotes the submatrix formed by the columns of
$\mathbf A$ with indices from $m$ to $n$}.  {Correspondingly,} the  training matrices  for  the  cluster $j$ in $\mathcal C_i$ {can be achieved from the} ${T_c}_j \times {T_c}_j$ DFT matrix $\mathbf  F_{c}$ as
\begin{align}
&\mathbf{T_c}_j^i =  \big[\mathbf  F_{c}^i  \big]_{1:M_j^i,}\label{train_center}
\end{align}
where $\mathbf F_{c}^i $ is defined as
\begin{align}
&\mathbf{F}_c^i =  \big[\mathbf  F_{c}  \big]_{1:\bar M^1,} \\
&\mathbf{F}_c^2 =  \big[\mathbf  F_{c}  \big]_{(\bar M^1+1):(\bar M^1+ \bar M^2),}\\
&\mathbf{F}_c^3 =  \big[\mathbf  F_{c}  \big]_{(\bar M^1+\bar M^2+1):(\bar M^1+ \bar M^2+\bar M^3).}
\end{align}
Then, the LS estimation of $\overline{\mathbf{H}}_{j}^{i}$  {can be formed as}
\begin{align}
\widehat {\overline{\mathbf{H}}}_{j\in \mathcal E}^{i}& =  \mathbf Y_{j \in \mathcal E} \{\mathbf {T_e}_j^{i}\}^H =   \mathbf{\overline{H}}_{j\in \mathcal E}^{i}
+\mathbf{N}_{ j }\{\mathbf {T_e}_j^{i}\}^H,\label{est_H_edge}\\
\widehat {\overline{\mathbf{H}}}_{j\in \mathcal C_i}^{i}& =  \mathbf Y_{j \in \mathcal C_i} \{\mathbf {T_c}_j^{i }\}^H =   \mathbf{\overline{H}}_{j \in \mathcal C_i}^{i}
+\mathbf{N}_{ j }\{\mathbf {T_c}_j^{i }\}^H,\label{est_H_center}
\end{align}
 {where the equations} \eqref{edge_tr_condition}, \eqref{cent_tr_condition},
 {and the properties} $\mathbf{T_e}_{ j}^{i}( \mathbf{T_e}_{ j}^{i})^H = \mathbf I$, $\mathbf{T_c}_{ j}^{i}( \mathbf{T_c}_{ j}^{i})^H = \mathbf I$ are utilized.

\subsection{\textcolor{blue}{The Recovering of the Covariance  Matrices $\mathbf{K}_{j\in \mathcal C_i}$ }}

We give the procedures  for the estimation of
$\mathbf{K}_{j\in \mathcal C_i}$ {in this subsection}. Multiplying $\mathbf Y_{j \in \mathcal C_i}$ by $\{\mathbf {F}_c^{i'}\}^H$, we can  obtain
\begin{align}
\mathbf \Upsilon_{j\in \mathcal C_i}  & = 	\sum_{i'\!=1,i'\!\neq i}^{3} \mathbf Y_{j \in \mathcal C_i} \{\mathbf {F}_c^{i'}\}^H  \notag\\
&=  \sum_{i'\!=1,i'\!\neq i}^{3}  \mathbf N_j  \{\mathbf {F}_c^{i'}\}^H\notag\\
	&~~~+\sum_{i'\!=1,i'\neq i}^{3}   \sum_{j'\in \mathcal C_{i'}} \!\!\!\mathbf{H}_{j}^{i'} \mathbf{B}_{j'}^{i'}\!\left[\mathbf{I}_{M_j^i}, \!\mathbf 0_{M_j^i \!\times (\bar M^i\!- \!M_j^i)  }\right],
\end{align}
\textcolor{blue}{where the facts that $\mathbf{T_c}_{j}^{i} \mathbf {F}_c^{i'} = \mathbf 0$
and  $\mathbf{T_c}_{j}^{i} \mathbf {F}_c^{i} = [\mathbf{I}_{M_j^i}, \mathbf 0_{M_j^i \times (\bar M^i- M_j^i)  }]$ are utilized in the above derivation.}

 \textcolor{blue}{Resorting to the properties $\mathbf N_j (\mathbf N_j)^H = \sum_{t=1}^{{T_c}_j} {{\mathbf n_j}_{t}} {\mathbf n_j}_{t}^H \approx {T_c}_j \mathbf I$, and $\left[\mathbf{I}_{M_j^i}, \!\mathbf 0_{M_j^i \!\times (\bar M^i\!- \!M_j^i)  }\right] \left[\mathbf{I}_{M_j^i}, \!\mathbf 0_{M_j^i \!\times (\bar M^i\!- \!M_j^i)  }\right]^H =  \mathbf{I}_{M_j^i}$, we can derive}
\begin{align}
&\mathbf \Upsilon_{j} (\mathbf \Upsilon_{j})^H  \notag\\
& \!\!\approx \!\!2 {T_c}_j \mathbf I+ \underbrace{\!\! { \left( \sum_{i'\!=1,i'\!\neq i}^{3}    \sum_{j'\in \mathcal C_{i'}} \!\!\!\mathbf{H}_{j}^{i'} \mathbf{B}_{j'}^{i'}\right)\!\!\!\!
	\left(  \sum_{i'\!=1,i'\!\neq i}^{3}   \sum_{j'\in \mathcal C_{i'}} \!\!\!\mathbf{H}_{j}^{i'} \mathbf{B}_{j'}^{i'}\right) ^H}\!\!}_{\mathbf \Sigma_j}\notag.
\end{align}
where ${\mathbf \Sigma_j}$ is the covariance matrix  for the
sum of all  the  interference channels.
Since the  power of each data stream  for  the
cell-center cluster is $p_{cent}$, the covariance matrix of  the
equivalent noise $\underline{\mathbf n}_i$  in data transmission stage  can  be given by
\begin{align}
\widehat{ \mathbf{K}}_{ j}  =
\mathbf{I} &+p_{cent} \mathbf \Sigma_j \notag\\
= \mathbf{I} &+p_{cent}\big(\mathbf \Upsilon_{j} (\mathbf \Upsilon_{j})^H  - 2 {T_c}_j \mathbf I\big) \label{est_K}{\color{blue}.}
\end{align}
\begin{remark}
The proposed channel estimation scheme for IA-SSR requires low dimensional training matrices. Hence, \textcolor{blue}{our scheme eliminates the the pilot contamination \cite{rusek2013scaling}, which hinders  the performance of multi-cell massive MIMO systems}.
\end{remark}

\section{ The Discussion about the Implementation of IA-SSR}

\subsection{The Overhead Analysis}

In  the  IA-SSR scheme,  {each cell-center cluster and
	each cell-edge cluster needs to feedback
	$M_j^i K_j  N_r$ and $\sum\limits_{i=1}^{3}M_j^i  K_j N_r$, respectively.}
 {Let us assume} that each complex channel coefficient is quantized into $Q$ bits, the  channel coherent block length is $T$, and the rate of the feedback channel is $F$ bits per symbol.
Taking into consideration of the overhead of \textcolor{blue}{both the training and the feedback, we can separately derive the effective sum-rates for the cell-center cluster and the cell-edge cluster as}
\begin{align} \label{eff-rate-cent}
\mathbb R_{j\in \mathcal C_i} = \alpha_{j\in \mathcal C_i}  \mathbb C_{j\in \mathcal C_i},
~~~~~\mathbb R_{j\in \mathcal E}\! = \alpha_{j\in \mathcal E}  \mathbb C_{j\in \mathcal E},
\end{align}
where
\begin{align}
\alpha_{j\in \mathcal C_i}  &= \max \left\lbrace 1-\frac{ {T_c}_j}{T}-\frac{M_j^i K_j  N_rQ}{FT},0 \right\rbrace   , \\
\alpha_{j\in \mathcal E} &= \max \left\lbrace 1-\frac{ {T_e}_j}{T}-\frac{\sum_{i=1}^{3}M_j^i  K_j N_r Q}{FT},0\right\rbrace .
\end{align}

\subsection{Cluster division for   IA-SSR}
\textcolor{blue}{
	In the proposed transmission scheme, it is  critical to divide the user clusters into  the cell-­edge clusters and cell-­center cluster.
	To   achieve a better performance, we would like to develop a  adaptive clustering method.
	Theoretically, for a specific cluster $j$, we can separately derive
	its achievable capacity under two cases. The first one is that the cluster $j$ belongs to the cell-edge area,
	and the second one is that the cluster $j$ lies in the cell-center area.
	The  capacity for the former case is denoted as $\mathbb C_{j\in \mathcal E}$ in \eqref{c_sum_cent}, while that for the latter one is
	$\mathbb C_{j\in \mathcal C_i}$ in \eqref{c_sum_edge}.
	Then, we can formulate the clustering method to maximize the achievable capacity of the whole system as
\begin{align}
\left\lbrace \begin{array}{ll}
j \in \mathcal E, ~~\text{if}~ \mathbb C_{j\in \mathcal E} > \mathbb C_{j\in \mathcal C_i}; \\
j \in \mathcal C_{i^*}, i^*= \arg \max\limits_i { \mathbb C_{j\in \mathcal C_i}}, ~~\text{otherwise}.
\end{array} \right.
\end{align}
	However, this scheme requires to estimate and feed
	back the  equivalent channels $\{ \underline{\mathbf{H}}_{j}^{i}\}$ between all the clusters and the three
	BSs, which would seriously degrade the effective system sum-rate.
	To deal with this challenges, we will develop  a low-overhead cluster division criterion.
}	
	
\textcolor{blue}{ 	
	Recalling  the optimal problem (P1),  we list some examples of the optimal $({S^1_j,S^2_j,S^3_j} )$ in Table \ref{table}.
	\begin{table}[!t]
	\centering
	\renewcommand{\arraystretch}{1.3}
	\caption{Optimal $ S^1_j , S^2_j , S^3_j $ with Given 	$M^1_j$ ,  $M^2_j$ , $M^3_j$ and $N_r$.}
	\label{table}
	\begin{tabular}{c|c|c|c}
	\hline
	\hline
	($M^1_j$ , $M^2_j$ , $M^3_j$ , $N_r$) & $\sum_{i=1}^{3}{S^i_j}$ &$(S^1_j , S^2_j ,  S^3_j )$& IA efficient?\\
	\hline
	(2 , 2 , 2 , 2) & 3&(1,1,1)&yes\\
	\hline
	(3 , 3 , 3 , 2) & 4&(2,1,1)&yes\\
	\hline
	(5 , 3 , 3 , 2) & 4&(2,1,1)& no\\
	\hline
	(4 , 4 , 4 , 4) & 6&(2,2,2)& yes\\
	\hline
	(5 , 4 , 4 , 4) & 6&(2,2,2)& yes\\
	\hline
	(7 , 4 , 4 , 4) & 6&(2,2,2)& no\\
	\hline
	\hline
	\end{tabular}	
	\end{table}
	As observed from Table \ref{table}, the {IA-based} cooperative transmission has  many advantages, but it is not
	always suitable for all  the clusters. Obviously, IA is efficient if the following condition holds,
	\begin{align}
	S^1_j + S^2_j +  S^3_j >\max( M^1_j , M^2_j ,  M^3_j),\label{IA_effectient2}
	\end{align}
	which means that the  IA-based  cooperative transmission  provides more data streams than  the  transmission  from  a  single BS.
	Otherwise, the cluster $j$ should be served  by $BS_i$ with maximal $M^i_j$ exclusively.
}

\textcolor{blue}{ 	
	Generally, IA is efficient if $M^1_j,M^2_j$ and $M^3_j$ is roughly same.
	According to Theorem 1, $M^i_j$ is closely related to
	the distance  from  $BS_i$  to the cluster $j$. With this result,  we can  achieve  that IA is efficient for the cell-edge cluster, which have similar distance from each BS.
	However,  $M^i_j$ is obviously greater than $M^{i'}_j$ for  the   cluster $j$ in  the  cell-center area $i$,  $i'\neq i$. In this case, $BS_i$   exclusively {provides} more data streams  for cluster $j$ than IA.
	{Based} on this observation, we can partition  the  clusters into sets $\mathcal{C}_1,\mathcal{C}_2,\mathcal{C}_3$ and $\mathcal{E}$    by following  criterion.
	\begin{align}
	\left\lbrace \begin{array}{ll}
	j \in \mathcal E,~~ \text{if }S^1_j + S^2_j +  S^3_j >\max( M^1_j , M^2_j ,  M^3_j);\\
	j \in \mathcal C_{i^*}, i^*= \arg \max\limits_i {M^i_j}, ~~\text{otherwise}.
	\end{array} \right.
	\end{align}
	Since the cell-edge cluster consumes much more training resources to acquire the CSI than the cell-center cluster, we can further modify the above criterion  through     taking into consideration the overhead as follows.
	\begin{theorem} \label{th22} For the IA-SSR scheme, the maximal effective spatial DoF is obtained by partitioning the clusters into sets $\mathcal{C}_1,\mathcal{C}_2,\mathcal{C}_3$ and $\mathcal{E}$ according to the following criterion:
	\begin{align}
	\left\lbrace \begin{array}{ll}
	\!\!\!\!\!j\! \in \!\mathcal E, \text{if }\alpha_{_{j\in \mathcal E}}  (S^1_j\! \!+ \!\!S^2_j\! \!+ \! \! S^3_j )\!\!>\!\!  \max(\!\alpha_{_{j\in \!\mathcal C_1}}\! M^1_j \!, \alpha_{_{j\in \!\mathcal C_2}}\!M^2_j ,  \!\alpha_{_{j\in \!\mathcal C_3}}\!M^3_j);\\
	\!\!\!\!\!j \in \mathcal C_{i^*}, i^*= \arg \max\limits_i(  \alpha_{_{j\in \mathcal C_i}}{M^i_j}), ~~\text{otherwise}.
	\end{array} \right.
	\end{align}
	\end {theorem}
	}

\subsection{ \textcolor{blue}{	 Extension    of IA-SSR to some general scenarios }}
\textcolor{blue}{In this subsection, we discuss some practical aspects about the implementation of the proposed transmission scheme.}

\begin{figure}[!t]
	\centering
	\includegraphics[width=85mm]{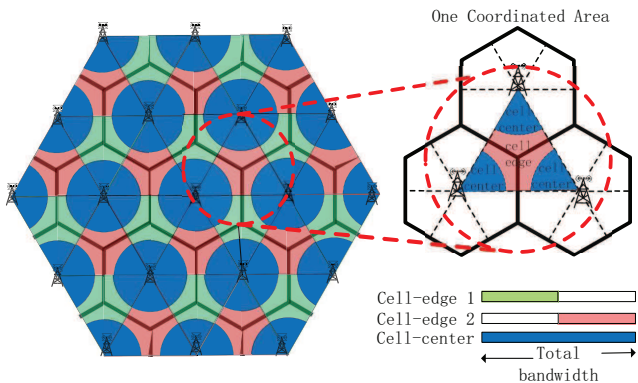}
	\caption{\textcolor{blue}{ Implementation of the proposed transmission scheme over the general multi-cell network.}}
	\label{fig:extend}
\end{figure}

\textcolor{blue} {\subsubsection{ Extension to the	general multi­cell networks}
In the previous sections, for the sake of clarity, we have focused on the core concepts of  the  proposed scheme in a three-cell scenario. Nevertheless,  the proposed scheme can be applied to the general multi-cell  networks. As shown in \figurename{ \ref{fig:extend}},
the whole coverage area can be partitioned into several  coordinated areas (CAs), and each CA   refers to the three adjacent sectors. To  avoid interference  among  CAs, we can  assign orthogonal  resources to the adjacent CAs. It can be checked from  \figurename{ \ref{fig:extend}} that only  two orthogonal  resources are enough. Therefore, we can equally divide the total bandwidth into two subbands and  separately  assign them  to the  cell-edge areas of  the  adjacent CAs. Since the distance between  two  cell-center areas is long  enough to ensure  the  large path loss, the mutual-interference  between different cell-center areas is low enough to be treated as noise. Such that   the cell-center users can use all the total bandwidth.
Under this frequency allocation scenario, the proposed transmission scheme can be  effectively  implemented in each CAs, and no coordination is required between the CAs.}
\begin{figure}[!t]
	\centering
	\includegraphics[width=85mm]{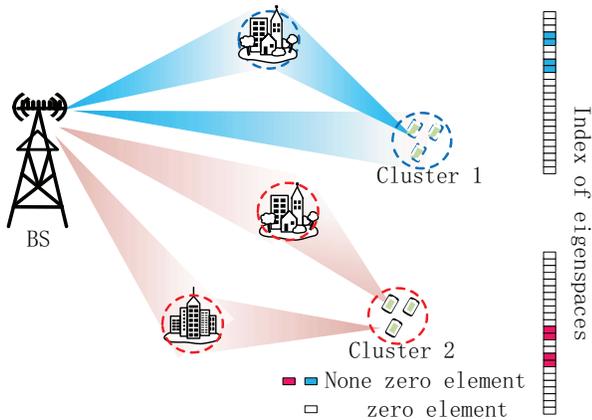}
	\caption{\textcolor{blue}{The scenario with multiple scattering rings.}}
	\label{fig:multi_ring}
\end{figure}
\textcolor{blue}{\subsubsection{ Beyond the one-ring model}
For mathematical convenience, the one-ring scattering model is used in
the previous sections. Nevertheless, the proposed  scheme can be  also applicable for the practical scenarios with multiple scattering rings, which is shown in \figurename{ \ref{fig:multi_ring}}.
 The main differences between the one-ring and the multiple-ring models
 can be summarized as follows.
 Under the multiple-ring model, the index set $\mathcal{I}_j^i$ is a composition of   multiple contiguous sequences, while that for the single-ring is just one  contiguous sequences.}

 \textcolor{blue}{\subsubsection{Implementation in TDD system}
	The proposed transmission scheme is proposed to reduce the overhead for the CSI acquiring in the FDD system. Nonetheless, the proposed scheme can be directly utilized for the TDD system without changing any components. Moreover, the proposed scheme can also gain advantage in the TDD system, such as improving the performance of the cell-edge clusters.
 }

\textcolor{blue}{For completeness, the  detailed IA-SSR  based cooperative transmission scheme is outlined in  Algorithm \ref{alg:2}.}
\begin{algorithm}[!t]
	\caption{ IA-SSR based Cooperative Transmission Scheme for  the  Multi-cell Massive MIMO system }
	\label{alg:2}
	\begin{algorithmic}[1]
		\STATE Initialize $\mathcal E = \emptyset$, $\mathcal C_i = \emptyset$.
		\FOR{$j=1:J$}
		
		\STATE Calculate $\mathbf B^i_j$ for  cluster $j$ by \eqref{B_edge}.
		\STATE Solve problem (P1).
		\textcolor{blue}{		\IF {$ \alpha_{_{j\in \mathcal E}}  (S^1_j\! \!+ \!\!S^2_j\! \!+ \! \! S^3_j ) > \max(\!\alpha_{_{j\in \!\mathcal C_1}}\! M^1_j \!, \alpha_{_{j\in \!\mathcal C_2}}\!M^2_j ,  \!\alpha_{_{j\in \!\mathcal C_3}}\!M^3_j)$}
			\STATE  $\mathcal E = \mathcal E \cup \{ j\} $
			\ELSE
			\STATE $ i= \arg \max_i(  \alpha_{_{j\in \mathcal C_i}}{M^i_j}), \mathcal C_i = \mathcal C_i \cup \{ j\} $
			\ENDIF}
		\ENDFOR
		\STATE Recalculate $\{\mathbf B^i_j\}$ for  clusters in  $\mathcal C_i$ by \eqref{B_center}.
		\STATE Design training matrices $ \{\mathbf{T_e}_j^i\}$  and $\{\mathbf{T_c}_j^i\}$
	    {with} \eqref{train_edge_1}-\eqref{train_center}.
		 {Estimate the} effective equivalent  channels $\{\mathbf H_j^i\}$
		and  {the channel covariances} $\{\mathbf K_j\}$  {with} \eqref{est_H_edge},\eqref{est_H_center} and \eqref{est_K}.
		\STATE \textcolor{blue}{ Calculate  the  inner precoder $\mathbf{V}_{j}^{i} $ for  the  cell-center cluster  with  \eqref{V_cellcenter}.}
		\STATE\textcolor{blue}{ Calculate  the inner precoder $\mathbf{V}_{j}^{i} $ for  the 	cell-edge cluster by solving the equations set \eqref{v1}-\eqref{v3}.}
		\STATE \textcolor{blue}{Calculate  the  decoding matrix $\mathbf{U}_{j,i} $ for  the  cell-edge cluster  through  \eqref{cal_U}.}
		\STATE Perform  the  power allocation  through  Algorithm \ref{alg:1}.
		\STATE \textcolor{blue}{Implement thedownlink data transmission with two-stage precoder $\mathbf P_{j}^i=\mathbf B_{j}^i\mathbf V_{j}^i$.}
	\end{algorithmic}
\end{algorithm}

\section{Numerical Results and Discussion}
In this section, we evaluate the proposed IA-SSR based cooperative transmission  scheme  through numerical simulations.
We consider a three-cell cellular system with  2 clusters in each cell-center area and 3 clusters in  { the} cell-edge area.  The number of users in each cluster is $3$.
The radius of the cell is 1 kilometer.
The distance between cell-center clusters and BS is 350 meters and the distance between cell-edge clusters and BS is 900 meters.
Each BS  possesses  $N_t = 128$ antennas, and each user has $N_r=2$ antennas. The  BS antenna  element  spacing  is  equal to  {the} half  wavelength.  The carrier frequency is 2GHz. We generate  the
massive MIMO channel according to \eqref{correlation} and \eqref{massive MIMO channel}. The free-space path loss (FSPL)  is considered. The variance of the noise is 1, and the signal-to-noise ratio (SNR)  is defined as  SNR $= \beta_{j\in \mathcal C_i}^i { P_{total} } $, which is the total transmit power normalized by  the  path loss of cell-center clusters.
The performance of  the  proposed IA-SSR scheme is compared with the directly extended two-stage precoding scheme  (DE scheme) for multi-cell massive MIMO network in \cite{muticell1} and the CoMP scheme with full CSIT in \cite{comp}.

\begin{figure}[!t]
\centering
\includegraphics[width=84mm]{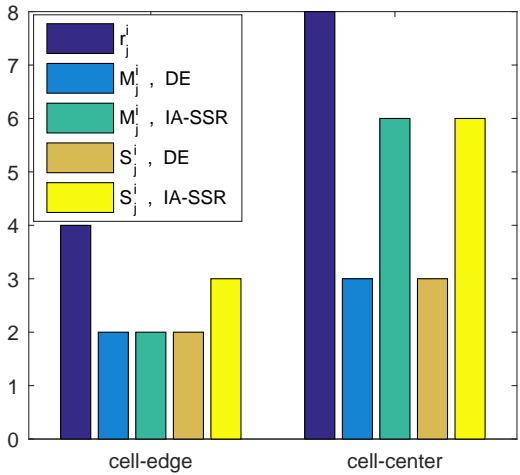}
\caption{Comparison of the rank of channel covariance matrix, the rank of effective equivalent  channel, and number of data streams in IA-SSR  scheme and that in DE scheme.}
\label{fig:dof}
\end{figure}

\figurename{ \ref{fig:dof}}  shows the  rank of
 the  spatial correlation matrix $r_j^i$, the rank of  the  effective equivalent  channel $M_j^i$, and the number of  data streams $S_j^i$ of  {both} DE and IA-SSR   {schemes}, respectively. In DE scheme, we get $M_{j\in \mathcal E}^i=2$ and $M_{j\in \mathcal C_i}^i=3$  , which are  much smaller than $r_{j\in \mathcal E}^i=4$ and $r_{j\in \mathcal C_i}^i=8$  due to the   AoD range overlap. In this case, $S_{j\in \mathcal E}^i=2$  is  smaller than the user number. As a result, one user is blocked. Whereas, $M_{j\in \mathcal C_i}^i = 6 $ in IA-SSR, which  doubles that in DE scheme. The $M_{j\in \mathcal E}^i$  remains unimproved in IA-SSR, which agrees with Remark 3. But IA improves the number $S_{j\in \mathcal E}^i$ of  data streams for  the  cell-edge cluster to 3, which is $3/2$ times of that of DE scheme.

\begin{figure}[!t]
	\centering
	\includegraphics[width=80mm]{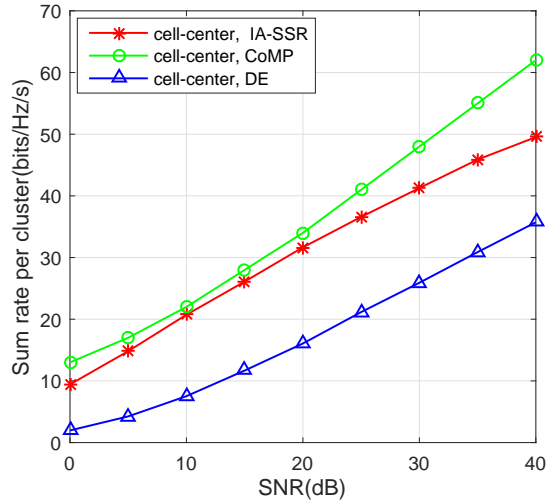}
	\caption{Sum-rate per cell-center cluster in IA-SSR scheme, DE scheme and CoMP scheme.}
	\label{fig:sum_rate_c}
\end{figure}
\begin{figure}[!t]
	\centering
	\includegraphics[width=80mm]{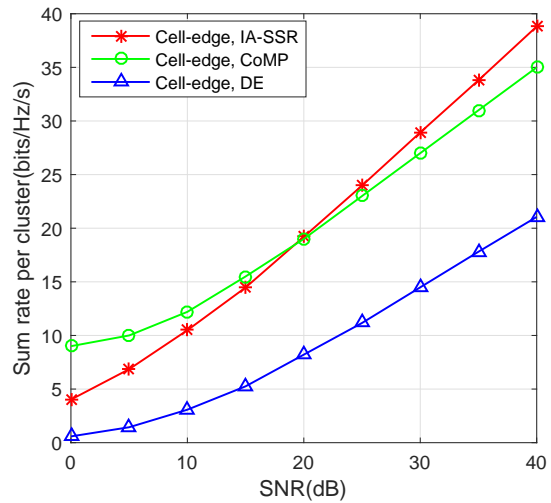}
	\caption{ Sum-rate per cell-edge cluster in IA-SSR scheme, DE scheme and CoMP scheme.}
	\label{fig:sum_rate_e}
\end{figure}

{\figurename{ \ref{fig:sum_rate_c}}/\figurename{ \ref{fig:sum_rate_e}}}  compares the sum-rate per cell-center/cell-edge cluster of  IA-SSR  scheme against that of DE scheme and CoMP scheme. The simulation results show that the sum-rate of  both the cell-center  and  cell-edge cluster in IA-SSR scheme is  obviously higher than that of DE scheme. The sum-rate of the cell-edge cluster is almost doubled. Whereas, the improvement for the cell-center cluster is not that significant. This phenomenon may  {look} strange as the number of data streams for the cell-center cluster is doubled, while that for the cell-edge cluster increases $3/2$ times. It can be explained as  follows: in IA-SSR scheme, the transmission power for the cell-center cluster is limited to avoid interference between different cell-centers. We can also see that the performance of the  proposed IA-SSR scheme is close to that of CoMP scheme with full CSIT. In high SNR regime, the sum-rate of cell-edge cluster in the IA-SSR scheme is even a little higher than that of CoMP scheme as IA provides more spatial DoF.



\begin{figure}[!t]
	\centering
	\includegraphics[width=74mm]{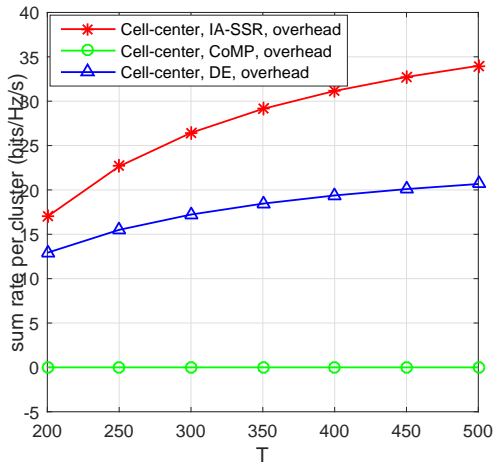}
	\caption{ Effective sum-rate per cell-center cluster versus channel coherent block length $T$ (SNR = 30dB, $F = 4, Q=16$).}
	\label{fig:overhead_c}
\end{figure}

\begin{figure}[!t]
	\centering
	\includegraphics[width=75mm]{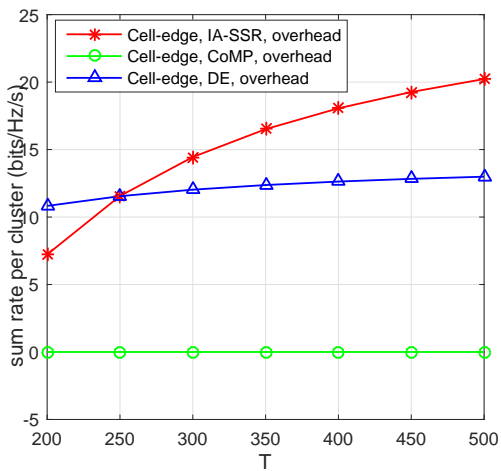}
	\caption{ Effective sum-rate per cell-edge cluster versus channel coherent block length $T$ (SNR = 30dB, $F = 4, Q=16$).}
	\label{fig:overhead_e}
\end{figure}

To uncover the  overhead  of our designed channel training scheme and CSI feedback, we present a numerical example of effective sum-rate in {\figurename{ \ref{fig:overhead_c}}}  and {\figurename{ \ref{fig:overhead_e}}}. The effective sum-rate of the IA-SSR scheme is computed according to \eqref{eff-rate-cent}, where SNR = 30dB, $F = 4$ and $ Q=16$ . The effective sum-rate of the DE scheme and CoMP scheme is computed in a similar approach.  From {\figurename{ \ref{fig:overhead_c}}} and {\figurename{ \ref{fig:overhead_e}},}  we can observe that the performance gap between IA-SSR  scheme  and DE scheme  {decreases} as $T$ decreases. The sum-rate of cell-center cluster in IA-SSR  scheme is always higher than that {in}  DE scheme. Whereas, the sum-rate of the cell-edge cluster in  IA-SSR  scheme becomes lower than that in DE scheme  when $T$ is  small. The {reason} behind this is that the amount of needed cell-edge clusters' CSI for IA is 3 times of that for DE scheme. We can also see that the effective sum-rate of CoMP scheme in considered $T$ is zero{, due to the fact} that in FDD mode all {the} channel resource is used for pilot transmission and full CSI feedback. Such that the CoMP scheme is only suited for TDD mode.

\begin{figure}[!t]
	\centering
	\includegraphics[width=84mm]{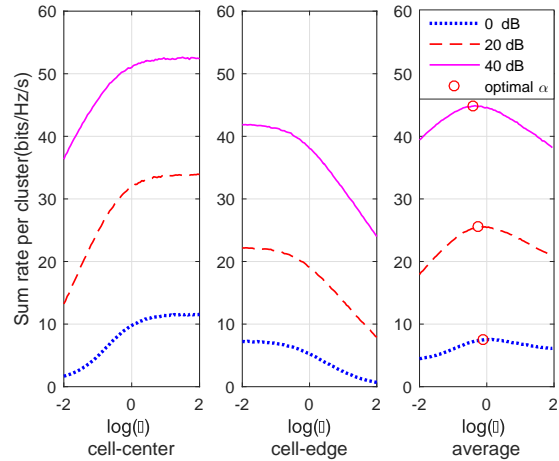}
	\caption{ Sum-rate per cluster of cell-center cluster, cell-edge cluster, and their average   versus $\log_{10}(\alpha)$ at SNR = 0dB, SNR = 20dB, and SNR = 40dB.}
	\label{fig:ratio}
\end{figure}

In order to evaluate  the proposed power allocation policy, we define the power splitting factor as
$$\alpha =  \frac{p_{cent}}{p_{edge}},$$
where $p_{edge}$ is the average  power allocated to each data stream for cell-edge cluster. {\figurename{ \ref{fig:ratio}}} presents the sum-rate per  cell-center cluster, the sum-rate per cell-edge cluster, and their average  versus the power splitting factor $\alpha$ (in $\log$ form), respectively. Three different SNRs, i.e., 0 dB{,} 20 dB, and 40 dB are adopted here. {\figurename{ \ref{fig:ratio}}} shows the same results with that we analyzed in section III.C. In small $\alpha$ regimes, the sum-rate of the cell-center cluster increases significantly  as  $\alpha$  grows, which results in   an  increase of the average sum-rate. Whereas, in large  $\alpha$ regimes, the  sum-rate   increase   of  the cell-center cluster is not obvious, but the sum-rate of the cell-edge cluster  decreases  seriously. As a result, the average  sum-rate decreases. Furthermore,  the optimal
$\alpha$ {(marked} by red circle)  obtained by our designed Algorithm \ref{alg:1}  perfectly matches    the peak point of average  sum-rate of simulation, which verifies the efficiency our power allocation algorithm. We can also see that the  optimal $\alpha$ is always smaller than 1. It means that the power of cell-center cluster is  limited within a lower level than that of cell-edge cluster. By comparing the  optimal $\alpha$ of three different SNRs, we realize that the optimal $\alpha$  decreases with the increase of SNR. The reason lies behind this is   the interference  dominates the interference plus noise at a high SNR  regime, which makes the sum-rate  of cell-center cluster is sensitive to interference. Therefore, the power for the cell-center cluster should be limited more strictly to avoid interference between different cell-centers at high SNR.

\begin{figure}[!t]
	\centering
	\includegraphics[width=78mm]{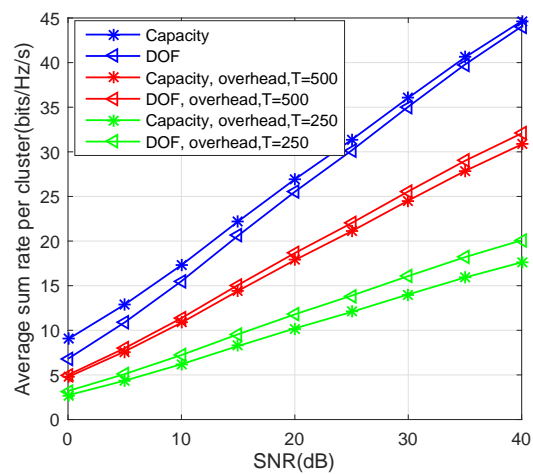}
	\caption{\textcolor{blue}{Comparison of the proposed effective DoF maximizing cluster division criterion and the  capacity maximizing criterion.}}
	\label{fig:division}
\end{figure}

\textcolor{blue}{
	\figurename{ \ref{fig:division}} compares the achievable sum rate  of the effective DoF maximizing criterion
with that of the achievable capacity maximizing criterion.
In this example, three clusters are randomly distributed in each cell.
From \figurename{ \ref{fig:division}}, we observe that, if
the overhead of CSI acquiring is  ignored,  the effective DoF maximizing criterion
results in a slight performance loss in the low and middle SNR regimes.
However,  if the  overhead is considered,   the effective sum-rate  for
the effective DoF maximizing criterion is higher than that of the capacity maximizing criterion, especially
under the scenario with small coherent time.
}

\begin{figure}[!t]
	\centering
	\includegraphics[width=81mm]{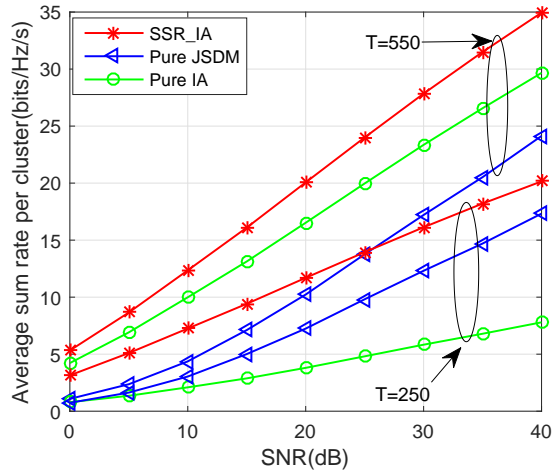}
	\caption{\textcolor{blue}{ A performance comparison of the proposed IA-SSR scheme, the pure JSDM scheme and pure IA scheme.}}
	\label{fig:IASSR_SSR_IA}
\end{figure}
\textcolor{blue}{
In order to better appreciate the gains of the proposed IA-SSR scheme, the effective sum-rate  of the proposed IA-SSR scheme is compared with those of the pure JSDM scheme and the pure IA scheme  in {\figurename{ \ref{fig:IASSR_SSR_IA}}}. The channel coherent block length $T$ is set as 250 and 550, respectively.   The results show that  the sum-rate of the pure JSDM scheme is higher than that of the pure IA scheme when $T$ is small. Contrarily, the pure IA scheme achieves better performance when $T$ is large. The reason behind this is that the IA scheme provides greater DoF than the JSDM scheme, but it consumes more time blocks to obtain global CSI.  Since the proposed IA-SSR scheme adaptively makes a tradeoff between the IA scheme and the JSDM scheme, it always outperforms  the pure IA scheme and the pure SSR scheme.
}

\begin{figure}[!t]
	\centering
	\includegraphics[width=75mm]{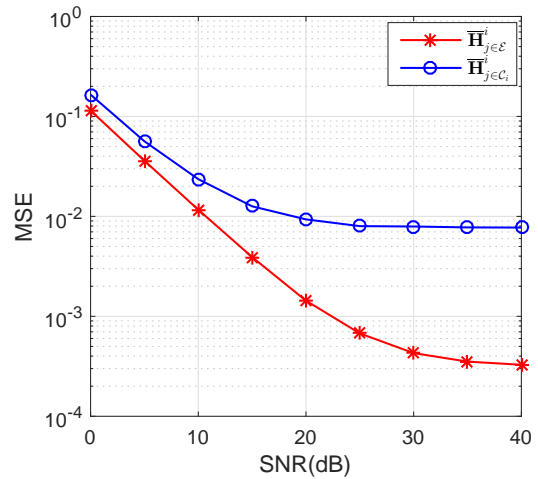}
	\caption{\textcolor{blue}{ MSE performance of the proposed the channel training scheme.}}
	\label{fig:channel_est}
\end{figure}

\textcolor{blue}{
	{\figurename{ \ref{fig:channel_est}}}  evaluates the performance of our proposed channel
	estimation scheme in terms of MSEs. It can be seen that the
	MSE for the cell-center clusters is higher than that of the cell-edge clusters. The reason behind this is the cell-center clusters
	suffer the interference from the other cell-center areas. With
	the SNR increasing, the MSEs for both the cell-center clusters
	and the cell-edge clusters have error floors. This phenomenon
	can be explained as follows. The complete orthogonality of
	channels for different users is impossible and the slight inter-cluster interference may arise when the practical number of
	antennas at BS are considered.
}

\section{ Conclusions }
In this paper, we investigated the two-stage precoding for  the  multi-cell  massive MIMO systems, where multi-antennas at user side  {was} considered. { First}, we  proposed an IA-SSR  based cooperative transmission scheme to efficiently  {apply} the two-stage precoding  {for} the multi-cell scenario. Then, the optimal  {power allocation} and  {the low overhead} channel training framework were  developed for  the  IA-SSR  scheme. Finally,  {the} numerical simulation { results} showed that the proposed IA-SSR scheme yields significant performance gain over  the  existing
methods.


%
%
\balance

\end{document}